\documentclass[twocolumn]{aastex631}
\received{December 10, 2021}
\revised{March 28, 2022; April 27, 2022}
\accepted{May 6, 2022}
\submitjournal{ApJ}

\shorttitle{Cloudlet Capture Model for TMC-1A}

\shortauthors{Hanawa, Sakai, \& Yamamoto}
\graphicspath{{./}{figures/}}

\begin{document}

\title{Cloudlet Capture Model for Asymmetric Molecular Emission
Lines Observed in TMC-1A with ALMA}

\correspondingauthor{Tomoyuki Hanawa}
\email{hanawa@faculty.chiba-u.jp}

\author{Tomoyuki Hanawa}
\affiliation{Center for Frontier Science, Chiba University, 
1-33 Yayoi-cho, Inage-ku, Chiba 263-8522, Japan}

\author{Nami Sakai}
\affiliation{RIKEN Cluster for Pioneering Research, 
2-1 Hirosawa, Wako-shi, Saitama 351-0198, Japan}

\author{Satoshi Yamamoto}
\affiliation{Department of Physics, The University of Tokyo, Bunkyo-ku, 
Tokyo 113-0033, Japan}



\begin{abstract}
TMC-1A is a protostellar source harboring a young protostar, 
IRAS 04365+2353, and shows a highly asymmetric features of 
a few 100 au scale in the molecular emission lines.
Blue-shifted emission is much stronger
in the CS ($J=5$-4) line than red-shifted one.
The asymmetry can be explained if the gas accretion is episodic 
and takes the form of cloudlet capture, 
given the cloudlet approached toward us. 
The gravity of the protostar transforms the cloudlet 
into a stream and changes its velocity along the flow. 
The emission from the cloudlet should be blue-shifted before
the periastron, while it should be red-shifted after the
periastron.  If a major part of cloudlet has not reached 
the periastron, the former should be dominant.
We perform hydrodynamical simulations to examine the
validity of the scenario.  Our numerical simulations
can reproduce the observed asymmetry if the orbit
of the cloudlet is inclined to the disk plane.
The inclination can explain the slow infall velocity
observed in the C$^{18}$O ($J$=2-1) line emission. Such  episodic accretion may occur in various  protostellar cores since actual clouds  could have inhomogeneous density  distribution.
We also discuss the implication of the cloudlet capture on observations of related objects. 
\end{abstract}



\section{Introduction \label{sec:intro}}

Young Stellar objects often show signatures of rotating disks 
and their formation is an integral part of star formation.
The disks are roughly symmetric 
around the rotation axis, though some of them have spirals and other
substructures \citep[see, e.g.,][]{andrews18,sakai19,nakatani20}.  
Thus, we often assume implicitly that they are almost symmetric
from the birth.  However, some young objects show highly asymmetric features according to
high-resolution molecular line observations with ALMA  \citep[see, e.g.,][for L1489, TMC-1A, Per-emb2 and GSS30-IRS5, respectively]{yen14,sakai16,pineda20,artur19}, even though
are not close binaries.
In addition, non-axisymmetric distribution of the infalling gas around the protostar is often suggested from the asymmetry of spectral line profiles, although it is spatially unresolved 
\cite[e.g., L483 and B335][respectively]{oya17,imai19}.
More recently, \cite{garufi22} have reported streamers in DG Tau and HL Tau.
Even for the Class II source GM Aur, the asymmetric infall of the gas of a remnant envelope or a cloud component has been reported \citep{huang21}.

Star forming gas clouds are turbulent and inhomogeneous \citep[see, e.g., the
review by][]{hennebelle12}. Hence, the gas
accretion onto young stellar objects can be variable 
on a short timescale, though the accretion rate is thought to decrease along the
evolution \citep{kueffmeier17}.
Based on the above recognition, \citet{dullemond19}
have proposed that the gas accretion is sporadic in the late 
stage of star formation and some Class II objects have 
secondary disks formed by the capture of a cloudlet 
\citep[see also][]{kueffmeier19}.  More recently, \cite{kueffmeier21} have
made numerical simulations in which the captured cloud forms an outer
disk surrounding a pre-existing inner disk. 

The argument by \cite{dullemond19} is supported by the
large arc-like feature observed in some sources such as
AB Aur \citep[see, e.g.,][]{fukagawa04}.  Their hydrodynamical simulations
reproduce the arc-like feature. Their model is an interesting
idea to be examined further, though the arc-like feature may be
produced by another mechanism such as Rossby wave instability \citep[see, e.g.,][]{miranda16}. 

Capture of a cloudlet may take place in an earlier stage of star formation.  
A few years earlier than \citet{dullemond19}, the asymmetric molecular distribution is
reported in the prestellar source, TMC-1A \citep{sakai16}.
TMC-1A, which harbors the Class I protostar IRAS 04365+2535, is a typical example showing such asymmetry in the molecular emission lines \citep[see, e.g.,][]{sakai16, bjerkelli16, harsono21}
and shows clear blue-red asymmetry in the molecular emission lines. 
Blue-shifted emission in the East part
is much stronger than red-shifted one in the West.  The morphology and degree of
asymmetry depend  on the chemical species emitting lines.
The highly asymmetric morphology is unlikely to be ascribed
to different chemical composition or excitation condition.  This asymmetry
may be short lived, since rotation around the protostar should reduce
the asymmetry on the local Keplerian timescale.

In order to explain observed features of TMC-1A, \cite{sakai16} introduced 
a ballistic model, in which only the gravity of the
protostar is assumed to act on an accreting gas element. We aim to
reexamine this picture by using hydrodynamic simulations in which
we take account of the collision of the infalling gas with
the disk.  
Another issue of  TMC-1A is the mass of the protostar. 
It is highly uncertain and ranges from 0.25~M$_\odot$
to 0.7~M$_\odot$ \citep{aso15,bjerkelli16,sakai16}. This uncertainty is in
part due to the difference in the molecular emission lines used for the mass
estimate. We should remember that the derived mass depends on the 
inclination of the gas motion to the line of sight.  The current estimate
is based on the assumption that the gas flow is confined in the 
disk plane.  This assumption may be invalid.

Our numerical simulations are similar to those of 
\citet{dullemond19,kueffmeier19,kueffmeier21} but updated in some respects.  
First, we take account of the presence of a rotating
disk around the protostar and its dynamical interaction with the infalling
gas.  Second, we take account of
warm gas surrounding the disk and protostar.  
The cloudlet and disk gas are cold and dense
while the pressure is the same as that of the surrounding
warm gas.  The temperature is assumed to remain nearly constant
at a few tens Kelvin and several hundreds Kelvin
in the cold and warm gases, respectively.
The pressure of the warm gas prevents the dispersal of 
a cloudlet seen in the isothermal model of \citet{kueffmeier19}.
Our model is similar to their adiabatic model
but the specific heat ratio is assumed to be $ \gamma = 1.05 $.
This low specific heat ratio mimics the short thermal timescale.
The temperature of the gas increases temporally through kinetic energy
dissipation by shock but goes back to its initial one 
on a timescale much shorter than the dynamical timescale.
The gas temperature increases little in our model 
since it is proportional $ T \propto \rho ^{\gamma -1} $.

Our hydrodynamic model can reproduce the blue-asymmetry of
the CS emission observed in TMC-1A 
under the assumption that the CS molecules are contained only
in the cloudlet.  This assumption is reasonable since the CS molecules
are often abundant in an infalling-rotating envelope but not in disks \citep{sakai14b}.
It can explain also the shift of SO emission peak to the disk center
\citep{sakai16}.
Furthermore, it can explain slow infall velocity observed in the
C$^{18}$O emission line \citep{aso21} if the orbital
plane of the cloudlet is nearly face-on.  

This paper is organized as follows. We describe our
model and numerical methods in \S 2, and results in \S 3. 
We compare our models with the observations in \S 4 and
discuss implications in \S 5.

\section{Model} 

\subsection{Basic Equations}

We solve the hydrodynamic equations on the cylindrical coordinates
according \cite{hanawa21}.  They have succeeded in conservation of the
the $ z $-component of the angular momentum and the free stream preservation.
The latter guarantees that it can solve a uniform flow without truncation
errors. See \cite{hanawa21} for more technical details including 
numerical tests.

We use the hydrodynamical equations,
\begin{eqnarray}
\frac{\partial \rho}{\partial t} + \mathbf{\nabla} 
\left( \rho \mathbf{v} \right) & = & 0, \label{mcon} \\
\frac{\partial}{\partial t} \left( \rho \mathbf{v} \right)
+ \mathbf{\nabla} \left( \rho \mathbf{vv} + P \mathbf{I} \right) & = & 
- \rho \mathbf{\nabla} \Phi , \label{Pcon} 
\end{eqnarray}
to describe gas accretion onto a protostar associated with a gas disk.
The symbols, $ \rho $ and $ P $, denote the density and pressure, respectively, 
while $\mathbf{v} $ and $ \Phi $ do the velocity and gravitational potential, 
respectively.  The gas pressure is expressed as
\begin{eqnarray}
P & = & \frac{k}{\bar{m}} \rho T , 
\end{eqnarray}
where $ T $, $ k $, and $ \bar{m} $ denote the temperature, 
Boltzmann constant, and mean molecular weight, respectively.

We assume that the gas consists of cold and warm components.
The disk and cloudlet are composed of
the cold molecular gas, while they are surrounded by a warm atomic gas.
We assume the mean molecular weight to be $2.3~m _{\rm H} $ 
and $1.27~m _{\rm H} $ for the cold and warm gases, respectively, 
where $ m _{\rm H} $ denotes the mass of a hydrogen atom.
The assumed mean molecular weight means that the cloud gas is
molecular while the warm gas is atomic and neutral.
Both the cold and warm gases maintain their temperatures since
their thermal timescales are much shorter than the dynamical one.
In order to mimic the nearly isothermal change, we assume that the
the gas is adiabatic with the specific heat ratio, $ \gamma = 1.05 $.
Then we can use the energy conservation
\begin{eqnarray}
\frac{\partial}{\partial t} \left(\rho E\right) + 
\mathbf{\nabla} \cdot \left[ \left( \rho E + P \right) \mathbf{v} \right]
 =  - \rho \mathbf{v} \cdot \mathbf{\nabla} \Phi , \label{Econ} \\
E  =  \frac{\mathbf{v}\cdot \mathbf{v}}{2} 
+ \frac{P}{\left(\gamma - 1 \right) \rho} ,
\end{eqnarray}
to follow the change in the pressure.

We introduce a color field, 
\begin{eqnarray}
c & = & 
\left\{ \begin{array}{ll}
1 & \mbox{(cloudlet)} \\
0 & \mbox{(warm gas)}\\
-1 & \mbox{(gas disk)} \\
\end{array} \right. .
\end{eqnarray}
to trace the cloudlet and gas disk. The color (scalar) field is traced by
\begin{eqnarray}
\frac{\partial}{\partial t} \left( c \rho \right)
+ \mathbf{\nabla} \cdot \left( c \rho \mathbf{v} \right)  =  0 .
\label{ctrace}
\end{eqnarray}

We use the cylindrical coordinates, $ (r, \varphi, z ) $, in our computation.
We locate the protostar at the origin, $ r = z = 0$.  We also use the
Cartesian coordinates, $ (x, y, z) = (r \cos \varphi, r \sin \varphi, z) $
in our presentation.  For simplicity,
we ignore the self-gravity of the gas.  This simplification
is justified since we consider a cloudlet of $ \la 10 ^{-4}~\mbox{M} _\odot$ 
and gas disk of $ < 10 ^{-2}~\mbox{M} _\odot$.
Then the gravitational potential is expressed as
\begin{eqnarray}
\Phi = 
\left\{ \begin{array}{lc} - \displaystyle \frac{GM}{\sqrt{r ^2 + z ^2}} 
& (r ^2 + z ^2 \ge a ^2 ) \\
- \displaystyle \frac{GM}{2 a ^3} \left( 3 a ^2 - r ^2 - z ^2\right) & 
(r ^2 + z ^2 < a ^2 ) \\ \end{array} \right. ,
\end{eqnarray}
where $ M $ and $ a $ denote the mass of the protostar and the length scale,
respectively. The mass of the protostar is uncertain and different values
are adopted in the literature \citep[0.68. 0.4, and 0.25 M$\odot$ in][respectively]
{aso15,bjerkelli16,sakai16}.  In this paper we adopt an intermediate value of 
$M = 0.53~\mbox{M} _\odot $. The velocity and timescale given in this paper are 
proportional to the square root of the mass ($ \propto M ^{1/2} $, while the
temperature is proportional to the mass ($ \propto M $).  We take the length
scale to be $ a = 50~\mbox{au}$ to avoid numerical difficulties due to 
strong gravity in the region of $ r \le 50~\mbox{au}$.

We solve Equations (\ref{mcon}), 
(\ref{Pcon}), (\ref{Econ}), and (\ref{ctrace}) simultaneously in our
simulations. We do not take account of magnetic fields, radiation 
processes, and turbulence for simplicity.

\subsection{Initial Model}

We assume that the warm gas is static and in an isothermal 
hydrostatic equilibrium.
\begin{eqnarray}
P & = & P _0 \exp \left( - \frac{\bar{m} _0 \Phi}{k T_0} \right) ,
\label{P0}
\end{eqnarray}
where $ P _0 $, $ T _0 $, and $ \bar{m} _0 $ denote the pressure in the region very
far from the star,
the initial temperature, and the mean molecular
weight, respectively. The pressure is set to be
$ P _0  = 1.56 \times 10 ^7~\mbox{K~cm} ^{-3} $ in this paper,
though it can be set scale-free.  
The solution is still valid if we multiply both the density
and pressure by the same arbitrary constant. 
The hydrogen is assumed to be in the atomic form 
in the warm gas and its mean molecular weight is evaluated to be
$ \bar{m} _0 = 1.27~m_{\rm H}$, where $ m _{\rm H} $ denotes the mass
of a hydrogen atom. Accordingly, the density is expressed as
\begin{eqnarray}
\rho & = & \frac{\bar{m} _0 P _0}{k T _0} 
\exp \left( - \frac{\bar{m} _0 \Phi}{k T_0} \right) .
\end{eqnarray}
We consider the density distribution where the density at $ r = a $ is 
enhanced from that at infinity by a factor $ e ^{5/3} = 5.29 $, i.e.,
$ k T _0 = (3/5) G M \bar{m} _0 / a  $.
This means that the assumed temperature is $ T _0 = 883~\mbox{K} $.
This temperature may be slightly higher than the real value, 
though it is highly uncertain \citep[see, e.g., 
Figure 1 of][]{dutrey14}.

If the temperature is lower, the warm gas is more concentrated around
the protostar but will not affect the result significantly because
the cloudlet is also compressed to have a higher density around the
protostar.

Next, we consider a gas disk rotating around the protostar.
The disk is assumed to be stationary and axially symmetric.
The rotation velocity is expressed as
\begin{eqnarray}
\mathbf{v} & = &
\left\{
\begin{array}{cc} v _\varphi (r) \mathbf{e} _\varphi & | z | < z _s (r) \\
0 & | z | \ge z _s (r) 
\end{array} \right. , 
\end{eqnarray}
where $ z _s (r) $ and $ \mathbf{e} _\varphi $ denote 
the half thickness of the disk and the unit vector in the
$ \varphi $-direction, respectively.
Then the equation of motion reduces to
\begin{eqnarray}
- \frac{v _{\varphi} {}^2}{r} + \frac{1}{\rho} \frac{\partial P}{\partial r}
+ \frac{\partial \Phi}{\partial r} & = & 0 , \label{motion_r} \\
\frac{1}{\rho} \frac{\partial P}{\partial z} + \frac{\partial \Phi}{\partial z}
& = & 0 . \label{motion_z}
\end{eqnarray}

We assume that the temperature is uniform 
at $ T = T _{\rm d} $ inside the disk ($ |z | < z _s(r) $),
while it is at $ T = T _0 $ outside the disk. 
Then the solutions of Equations (\ref{motion_r}) and (\ref{motion_z}) 
are expressed as 
\begin{eqnarray} 
\rho (r, z) & = & \frac{\bar{m} _{\rm d}}{k T} P (r, z) , \\
P (r, z) & = & P _s (r) \exp \left\{ - \frac{\bar{m} _{\rm d} [\Phi (r, z) - \Phi (r, z _s)]}
{k T _{\rm d}} \right\} , \label{disk_P} \\
P _s (r) & = & P _0 \exp 
\left[ - \frac{\bar{m} _0 \Phi (r, z _s)}{k T _0} \right] . \\
v _\varphi ^2 & = & r \left( 1 - \frac{\bar{m} _0 T _{\rm d}}{\bar{m} _{\rm d} T _0} \right) 
\left( \frac{\partial \Phi _s}{\partial r} + \frac{d z _s}{dr} 
\frac{\partial \Phi _s}{\partial z} \right) ,
\end{eqnarray}
where $ \bar{m} _{\rm d} $ denotes the mean molecular weight of the disk gas and
is assumed to be $ \bar{m} _{\rm d} = 2.3~m _{\rm H} $.
Note that this model accomplishes the pressure balance on the disk surface.  

We assume that the disk has the initial radius, $ r _{\rm d} $, and the half thickness, 
\begin{eqnarray}
z _s & = & \beta \sqrt{r _{\rm d} {}^2 - r ^2} ,
\end{eqnarray}
where $ \beta $ is a non-dimensional free parameter and taken to be 0.2 in this paper.  
This model can not reproduce flaring of the outer disk, because the spatial resolution is limited. This
simple model is more durable against numerical instability than more realistic one 
due to stability against hydrodynamic waves.

We assume that a cloudlet occupies a spherical region,
\begin{eqnarray}
\left( x - x _{\rm c} \right) ^2 + \left(y - y _{\rm c} \right) ^2
+ \left( z - z _{\rm c} \right) ^2 \le  a _c {} ^2 , \\
\mathbf{r} _c = \left( \begin{array}{c} x _{\rm c} \\ y _{\rm c} \\ z _{\rm c} \end{array} \right)
 =   \left( \begin{array}{c} r _{\rm c} \cos \psi _{\rm c} \\ 0 \\
r _{\rm c} \sin \psi _{\rm c} \end{array} \right) , \label{cloudlet_init} 
\end{eqnarray}
at the initial stage, where $ r  _{\rm c} $ , $ a _{\rm c} $, and $ \psi _{\rm c} $ denote
the distance  to the cloud center, the cloudlet radius, and
the inclination of the orbit of the cloudlet to the disk, respectively. 
The initial
pressure inside the cloudlet is the same as given by Equation (\ref{P0}).
We assume that the cloudlet has the initial temperature, $ T _{\rm c} $,
and the mean molecular weight, $ \bar{m} _{\rm c} = 2.3~m _{\rm H} $. 
Then, the density is expressed as
\begin{eqnarray}
\rho & = & \frac{\bar{m} _c P _0}{k T _c} 
\exp \left( - \frac{\bar{m} _0 \Phi}{k T_0} \right) .
\label{density0}
\end{eqnarray}
The initial velocity is uniform at
\begin{eqnarray}
\mathbf{v} _{\rm c} & = &
\left( \begin{array}{c}
v _{c,x} \\
v _{c,y} \\
v _{c,z}
\end{array} \right) =
\left( \begin{array}{c}
- \cos \psi _{\rm c} \sqrt{\displaystyle \frac{2GM}{r _{\rm c}} - \frac{2 GM}{r _{\min}}} \\
\displaystyle \frac{\sqrt{ 2 G M r _{\min}}}{r _{\rm c}} \\
- \sin \psi  _{\rm c}  \sqrt{\displaystyle \frac{2GM}{r _{\rm c}} - \frac{2 GM}{r _{\min}}} 
\end{array} \right)
\end{eqnarray}
inside the cloudlet. This initial velocity coincides with the velocity of 
a particle having a parabolic orbit with the periastron, $ r _{\rm min} $,
at the distance $ r _{\rm c} $.  The parabolic orbit lies in the $xz$ plane
and is inclined by $ \psi _c$ from the $ z $-axis.

The temperature is assumed to be $ T _{\rm c} = 0.015 G M \bar{m} _c / k a $
in the cloudlet and $ T _{\rm d} = 0.03 G M \bar{m} _d /k a $ in the disk.
Accordingly it is $ T _{\rm c} = 39~\mbox{K} $ in the cloudlet and $ T _{\rm d} = 78~\mbox{K} $
in the disk, 
for $ M = 0.53~\mbox{M} _\odot $ and $ \bar{m} _{\rm c} = \bar{m} _{\rm d} = 2.3 m _{\rm H}$.

The numerical grid is designed so that the radial spatial resolution
is constant at $ \Delta r = 1~\mbox{au} $ and each numerical cell is
almost isotropic, $ r \Delta \varphi \simeq \Delta r $,
in the inner region of $ r \le 64~\mbox{au} $.
In the outer region of $ r > 64~\mbox{au}  $, the  angular resolution 
is constant at $ \Delta \varphi = 0.938^\circ $  and the radial
spatial resolution is $ \Delta r = 1.56 \times 10 ^{-2} r $.
The vertical spatial
resolution is $ \Delta z = 1~\mbox{au} $ in near the mid plane
($ | z | < 64.5~\mbox{au} $ and $ \Delta z = 1.56 \times 10 ^{-2} |z| $ 
in the outer regions of $ | z | > 64.5~\mbox{au} $.  The numerical cell 
covers the cylindrical region of $ r _{\rm out} $ and $ |z | < z _{\rm out} $.

Table \ref{tab:model} summarizes the models shown in the 
following sections.  The cloudlet is 500 au away from the central
star at the initial stage except for in model A$^\prime$.  
The orbit of the cloudlet is coplanar to the disk in models 
A, A$^\prime$, and D, while it is inclined by 30$^\circ$ in models
B and C.

\begin{table}[ht]
\caption{Model Parameters \label{tab:model}}
   \begin{center}
    \begin{tabular}{cccccc}
    \hline
  model &  $ r _{\rm out}{}^a $&  $ z _{\rm out}{}^b $ & $r _{\rm c} {}^c $ 
 & $ a _{\rm c} {}^d$  & $ \psi _{\rm c} {}^e $  \\
 \hline
 A & ~624 au & 246 au & 500 au & 100 au
 & $ ~0^\circ$  \\
 A$^\prime$ &  1254 au &  246 au &  1000 au & 100 au & $ ~0^\circ$  \\
 B & ~624 au &  392 au &  ~500 au & 100 au & $ 30^\circ$  \\
 C & ~775 au & 535 au & ~500 au & 200 au & $ 30^\circ$ \\
 D & ~775 au  &  336 au & ~500 au & 200 au &
    $ ~0^\circ$ \\
  \hline
    \end{tabular}
    \end{center}
$ ^a $~The outer radius of the computation domain. \\
$ ^b $~The half height of the computation domain. \\
$ ^c $~The initial radial distance of the cloudlet. \\
$ ^d $~The initial radius of the cloudlet. \\
$ ^e $~The inclination of the cloudlet orbital plane to the disk plane. 
See Equation (\ref{cloudlet_init}) for more details.
\end{table}

\subsection{Mock Observation}

We evaluate the emission expected from our numerical simulations as a post process
for comparison with molecular line emission observed.  We assume that
our line of sight is parallel to 
\begin{eqnarray}
\mathbf{n} _3 & = & \left( \begin{array}{c} \sin \theta _{\rm obs} \cos \varphi _{\rm obs} \\
\sin \theta _{\rm obs} \sin \varphi _{\rm obs} \\ \cos \theta _{\rm obs} \end{array} \right) ,
\end{eqnarray}
where $ \theta _{\rm obs} $ and $ \varphi _{\rm obs} $ specify the
location of the observer in the spherical coordinates. 
The disk inclination angle is given by $ i = 180 ^\circ - \theta _{\rm obs} $.
The radial velocity is evaluated to be
$ V = \mathbf{v} \cdot \mathbf{n} _3 $ while the observer is located
in the direction of $ - \mathbf{n} _3 $. Using the unit vectors,
\begin{eqnarray}
\mathbf{n} _1 & = & \left( 
\begin{array}{c} \sin \chi _{\rm obs} \sin \varphi _{\rm obs} 
- \cos \chi _{\rm obs} \cos \theta _{\rm obs} \cos \varphi _{\rm obs} \\
- \sin \chi _{\rm obs} \cos \varphi _{\rm obs} 
- \cos \chi  _{\rm obs} \cos \theta _{\rm obs} \sin \varphi _{\rm obs} \\
\cos \chi _{\rm obs} \sin \theta _{\rm obs} \end{array} \right) , \nonumber \\
\\
\mathbf{n} _2 & = & \left( \begin{array}{c} 
- \cos \chi _{\rm obs} \sin \varphi _{\rm obs} - \sin \chi _{\rm obs} 
\cos \theta _{\rm obs} \cos \varphi _{\rm obs} \\
\cos \chi _{\rm obs} \sin \varphi _{\rm obs}
- \sin \chi _{\rm obs} \cos \theta \sin \varphi _{\rm obs} \\
- \cos \chi _{\rm obs} \sin \theta _{\rm obs} \end{array} \right), \nonumber \\
\end{eqnarray}
where $ \chi _{\rm obs} $ denotes the disk position angle on the sky.
We define the position angle so that it increases counter-clockwise from the North. Then,
we can define the Cartesian coordinates,
\begin{eqnarray}
\mathbf{r} & = & X \mathbf{n} _1 + Y \mathbf{n} _2 + s \mathbf{n} _3 ,
\end{eqnarray}
where $ X $ and $ Y $ denote the projected distance from the protostar to the West 
and that to the North, respectively.

Our evaluation is based on the the simple assumption that the opacity at
the line center is the same in the whole region.  In other words, we 
ignore possible variation in the excitation temperature and abundance.  
Then, the optical depth is evaluated to be 
\begin{eqnarray}
\tau (X,Y, V ) & = & \kappa _0 \Sigma (X, Y, V) , \\
\Sigma (X, Y, V) & = & \int _{c>0} \frac{c (\mathbf{r}) \rho (\mathbf{r})}{\sqrt{2\pi} \sigma} \nonumber \\
& & \exp 
\left\{ - \left[ \frac{\mathbf{v} (\mathbf{r}) \cdot \mathbf{n} _3 - V }{2 \sigma ^2} 
\right]^2 \right\} ds , \label{sigma-los} \\
\mathbf{r} & = & X \mathbf{n} _1 + Y \mathbf{n} _2 + s \mathbf{n} _3 ,
\end{eqnarray}
where $ V $, $ \kappa_ 0 $, and $ \sigma $ denote the radial velocity, 
the opacity at the line center, and velocity dispersion, respectively. 
We assume $ \sigma = 0.153~\mbox{km~s}^{-1} $
to be slightly smaller than the velocity resolution of the observation 
($ 0.4~\mbox{km~s}^{-1} $)
so that we obtain smooth channel maps.
The intensity is evaluated to be
\begin{eqnarray}
I (X, Y, V) & = & I _0 \{ 1 - \exp [- \tau (X, Y, V) ] \} ,  \label{RTeq}
\end{eqnarray}
where $ I _0 $ denotes the intensity at the saturation level
and should coincide with the Planck function at the 
excitation temperature of the line emitting molecule.
When the line is optically thin, Equation (\ref{RTeq}) reduces to
\begin{eqnarray}
I (X, Y, V) & = & I _0 \kappa _0 \Sigma (X, Y, V) . 
\end{eqnarray}
Thus, we compare the column density per unit velocity, 
$ \Sigma (X, Y, V) $,
with the observed intensity, since our numerical model cannot
evaluate the opacity and excitation temperature quantitatively.
The integrated intensity should be proportional to the column density along
the line of sight as far as the line is optically thin and the temperature
is uniform.
We should evaluate the parameters, $ \kappa _0 $, $ \sigma $, and $ I _0 $ 
from the abundance and excitation temperature for each transition.  
However we do not take account of the physical processes in our numerical
simulation and hence the derived temperature is not accurate enough.
In this paper, we assume $ \sigma = 0.153 $~km~s$^{-1}$ and take
other constants arbitrarily.

\begin{figure*}[ht]
\plottwo{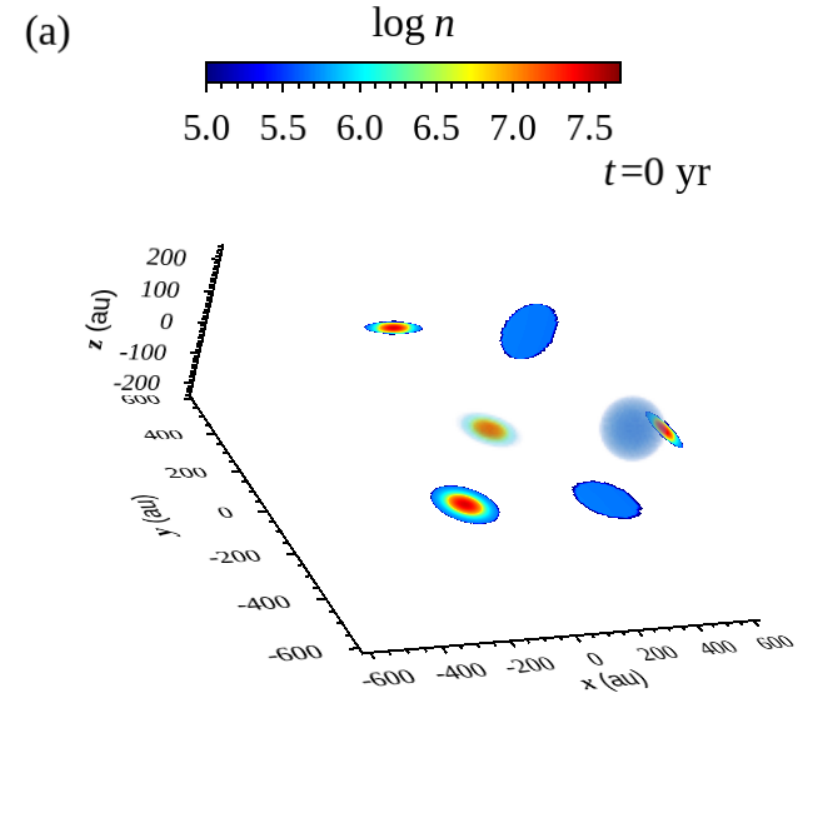}{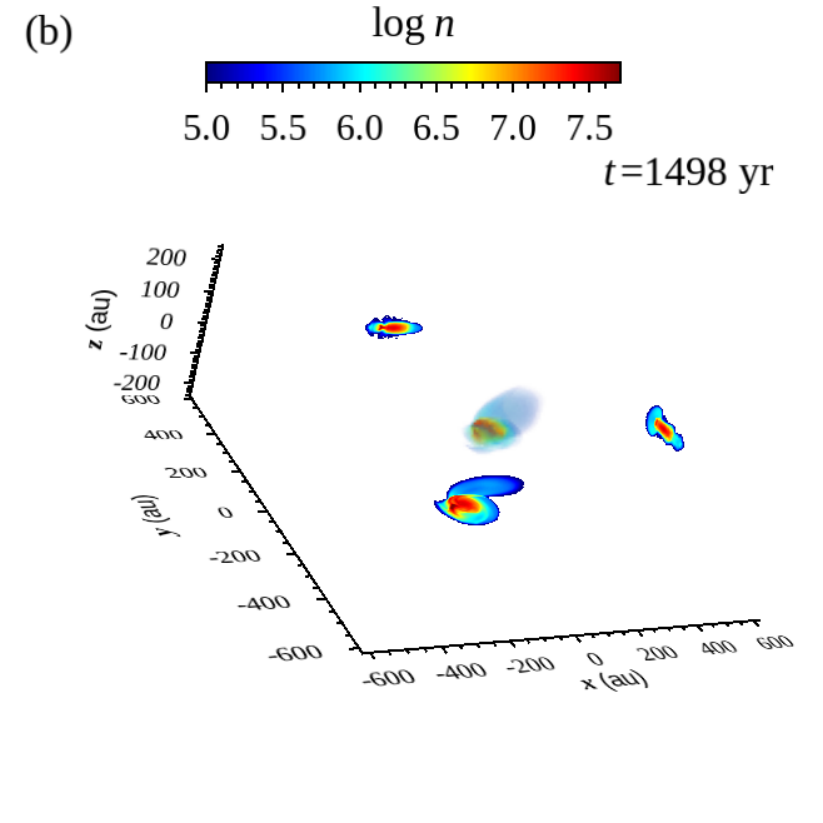}
\caption{Panels a and b show the initial state ($ t = 0$) and the stage at
$t = 1.498 \times 10 ^3 $~yr, respectively 
by combination of the volume rendering
and the cross sections.  A spherical cloudlet is located at
a distance of 500 au from the star and disk at the 
initial stage.\label{fig-init}}
\end{figure*}
\section{Results}

\subsection{Model A}

First we examine a prototypical model named A, 
in which the cloudlet
has the radius, $ a _{\rm c} = 100~\mbox{au} $ and is located at $ r _{\rm c} = 500~\mbox{au}$
at the initial stage, $ t = 0~\mbox{yr} $. The mass and average density of the cloudlet are
$ M _{\rm cl} = 1.27 \times 10 ^{-5}~\mbox{M} _\odot $ and 
$ \bar{\rho} _{\rm cl} = 1.80 \times 10 ^{-18}~\mbox{g~cm}^{-3} $, respectively.
The latter corresponds to the average number density, 
$ \bar{n} _{\rm c;} = 4.73 \times 10 ^5~\mbox{cm}^{-3} $.
The initial velocity of the cloudlet is 
1.37~km~s$^{-1}$. The disk half thickness is set to be 20~au $(\beta = 0.2)$.

Figure \ref{fig-init} shows the initial stage, $ t = 0 $, and an  early
stage of the collision of the cloudlet with the disk, $ t = 1498$~yr.
The head of the cloudlet touches the disk edge at $ t = 1130 $~yr.
The color denotes the density on the planes $ x = 0 $, $ y = 0$, and $ z = 0 $ in each panel.
The distribution of the cold gas is also shown by the volume rendering.
The color scale on the top of each panel is for the cross sections, although
a similar color scale is used for the volume rendering. 
The cloudlet shaves an outer part of the disk by the ram pressure.
When colliding, the density is a little higher in the disk than in
the cloudlet.  Still, the ram pressure exceeds the disk gas pressure
since the infall velocity is much higher than the sound speed.

\begin{figure}[ht]
    \epsscale{1.2}
    \plotone{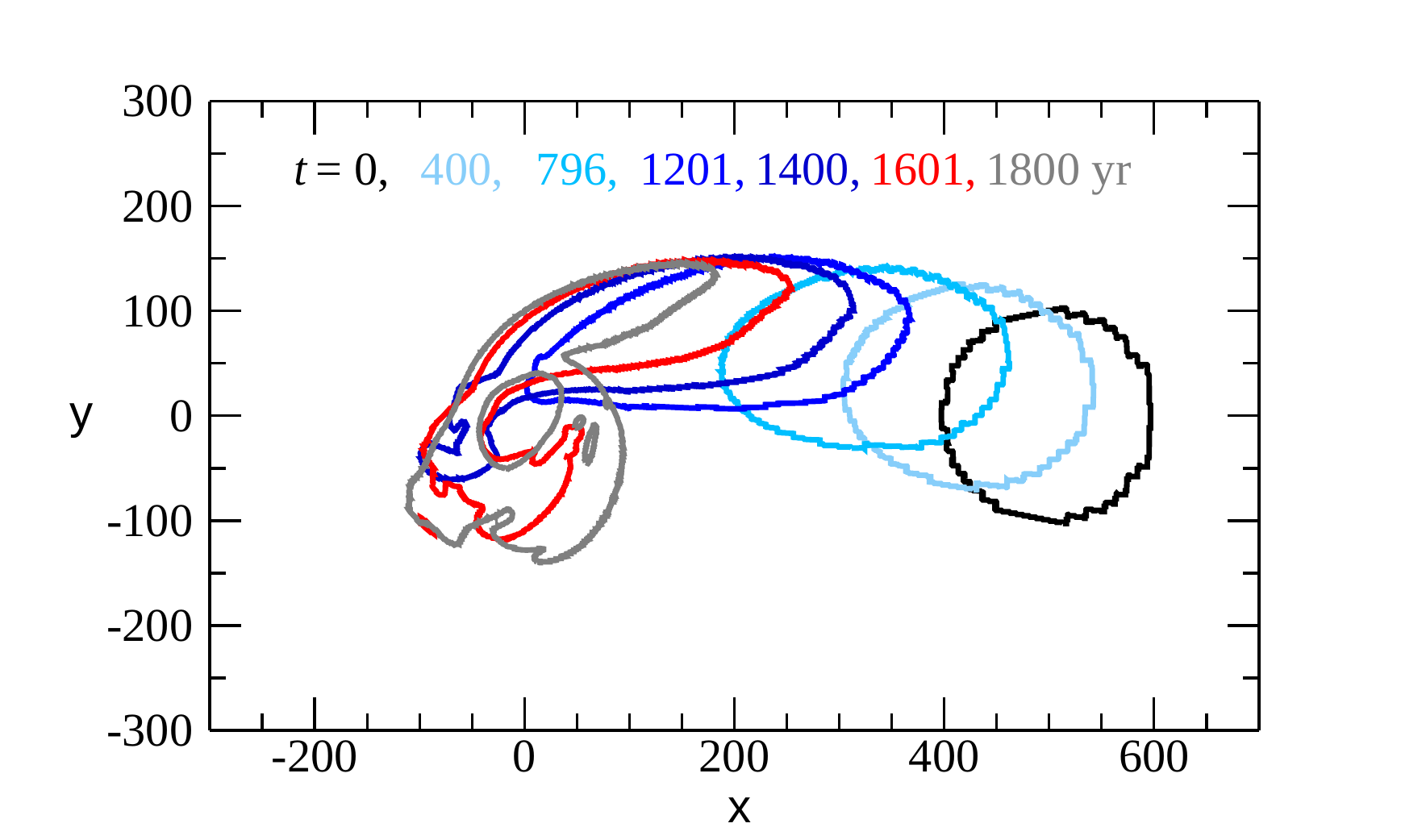}
    \caption{The evolution of the newly accreting
    cloudlet in model A. Each curve denotes the projection of the
    outer boundary of the accreting gas. The ordinate and abscissa
    denote $ x = r \cos \varphi$ and $ y = r \sin \varphi$, respectively.
    The epoch is shown by the same color on the top of the panel (Model A).}
    \label{fig:projected}
\end{figure}

Figure \ref{fig:projected} denotes the deformation of the cloudlet in the period,
$ t \le 1800 $~yr.  Each curve denotes the surface of the cloudlet projected on
the $ z = 0 $ plane.  We evaluated the cloudlet surface from the surface density of 
the cloudlet, 
\begin{eqnarray}
\Sigma _{\rm c} (x, y) & = & \int _{c>0} c (r, \varphi,
z) \rho (r, \varphi, z) dz , 
\end{eqnarray}
where $ (x, y)  = (r \cos \varphi, r \sin \varphi)$.
The tidal force of the protostar stretches the cloudlet in the direction
of infall; the front side is accelerated more than the rear side.  The cloudlet
is confined by the warm gas and does not expand appreciably. At $ t = 1498 $~yr,
the head of the cloudlet is compressed by the collision with the gas disk.
The compression increases the number density up to $ 6.24 \times 10 ^7~\mbox{cm}^{-3} $.
At $ t = 1734 $~yr, the disk is partly covered by the cloudlet. 
After the collision, the cloudlet changes its form; a part of it
accretes on the disk while the rest leaves the protostar.

\begin{figure*}
\plottwo{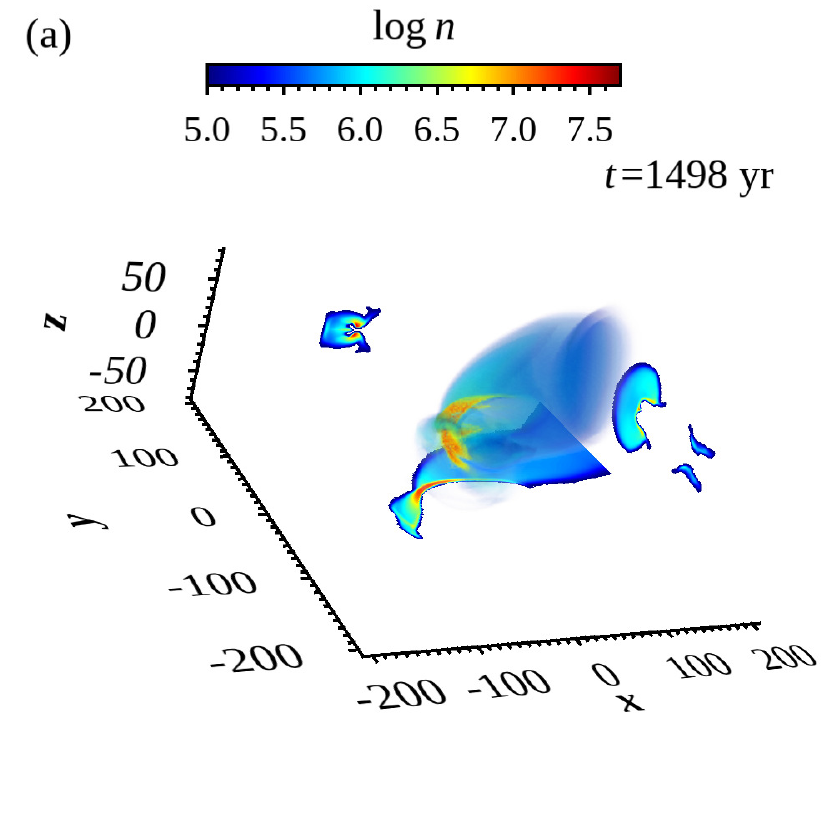}{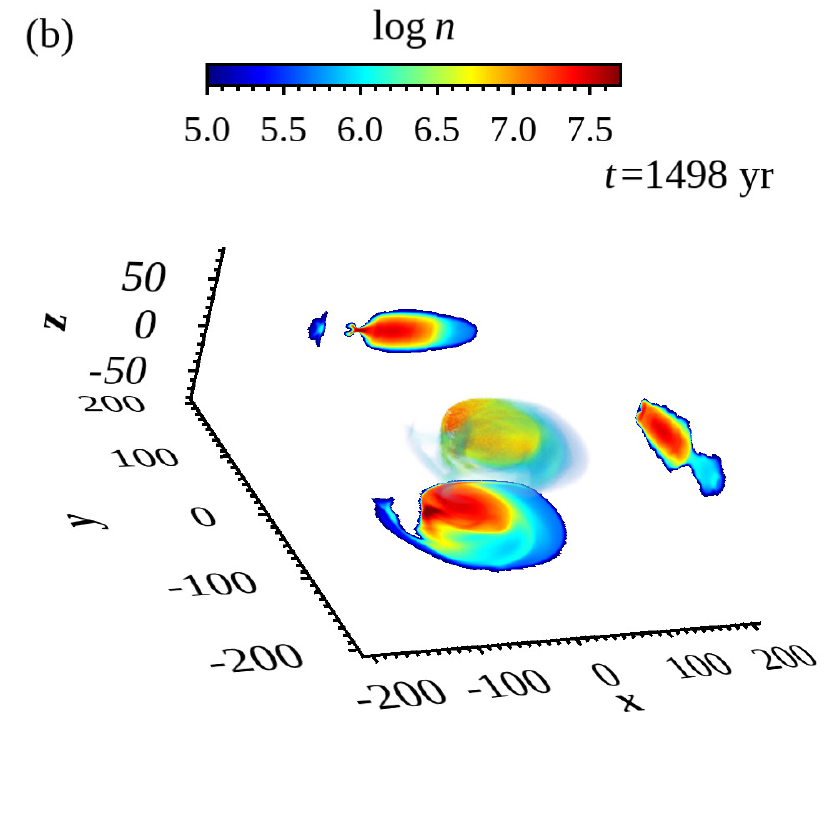}
\plottwo{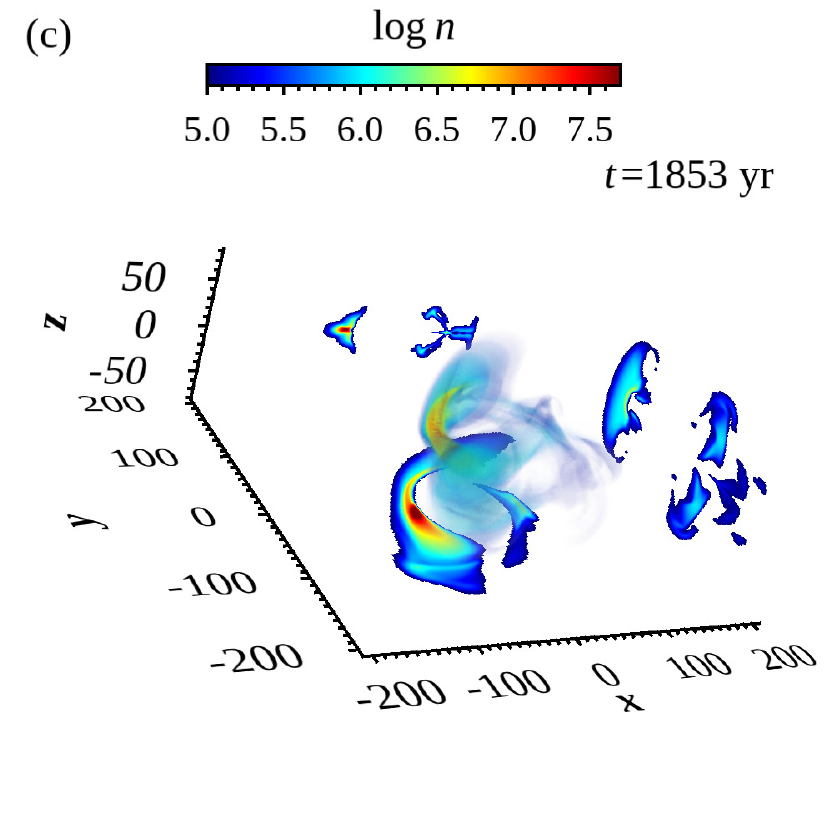}{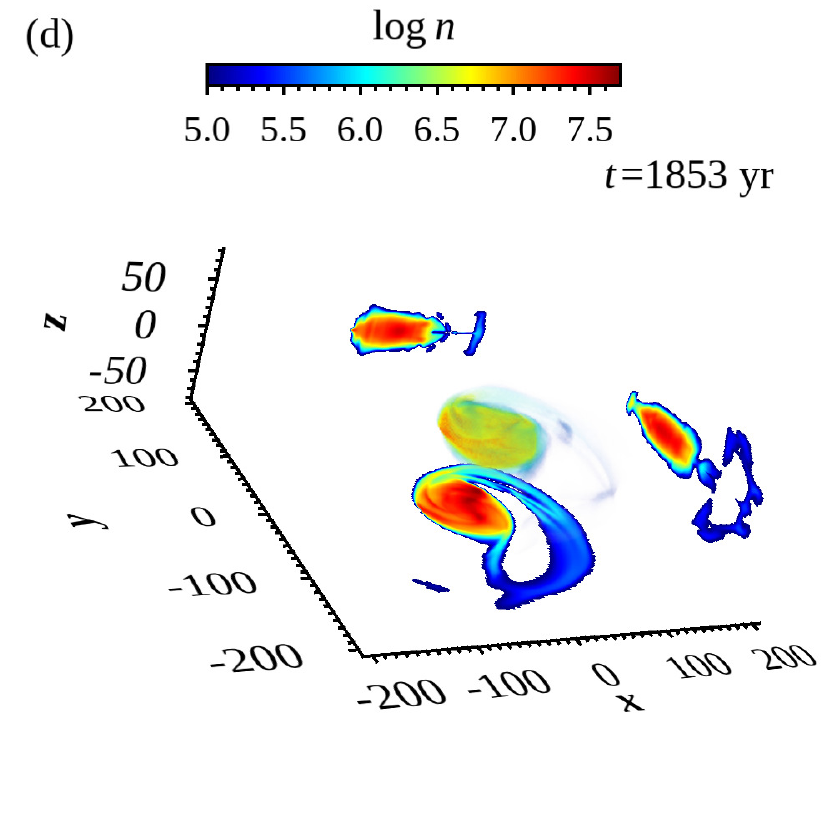}
\caption{Panels a and c denote the distorted cloudlet in model A while panels b and d do the disk.  The upper panels denote the stage
at $ t = 1498~\mbox{yr} $, while the lower ones do that at
$ t = 1853 $~yr. \label{EDA}}
\end{figure*}

The upper panels of Figure \ref{EDA} show 
the structure of the cloudlet and disk at
$ t = 1498 $~yr.  Figure \ref{EDA}a shows the gas of $ c > 0 $ (cloudlet)
while Figure \ref{EDA}b does that of $ c < 0 $ (disk). The cloudlet is
also bored by the disk to be separated into upper and lower halves.
Above and below the disk surface, the cloudlet gas continues to infall,
while the infall is blocked by the disk near the midplane.  The inner edge
of the cloudlet is shock compressed on the disk outer edge. 
On the other hand, the corresponding part of the disk is 
shaved to form an arm and accordingly the disk is highly asymmetric.

The lower panels of Figure \ref{EDA} show the cloudlet and
disk at $ t = 1853~\mbox{yr} $. 
Main part of the former cloudlet covers a substantial fraction of the the disk and the other small fraction is scattered outward.
The disk has several trailing arms in the outer region. The arms are shock waves induced by the impact of the collision with the cloudlet.
The inner disk is also appreciably affected, though the details are subject to change.
Note that the gravity is artificially softened in the region of 
$ r < 50~\mbox{au} $ in our model and accordingly the gas motion 
is not reliable there.

Figure~\ref{fig:projected} shows the surface
density distributions in the late stages. Most of the former cloudlet rotates around the protostar 
to accrete onto the disk, while a part of it forms an arm extending outward to be
ejected.  This ejection may be an artifact of our modeling in which we did not
take account of angular momentum extraction by magnetic force or energy dissipation
through radiation.  However, it is plausible that a part of infalling gas is ejected
from the protostar unless the energy dissipation is not efficient.

\subsection{Model A$^\prime$}

\begin{figure}[ht]
\epsscale{1.2}
    \plotone{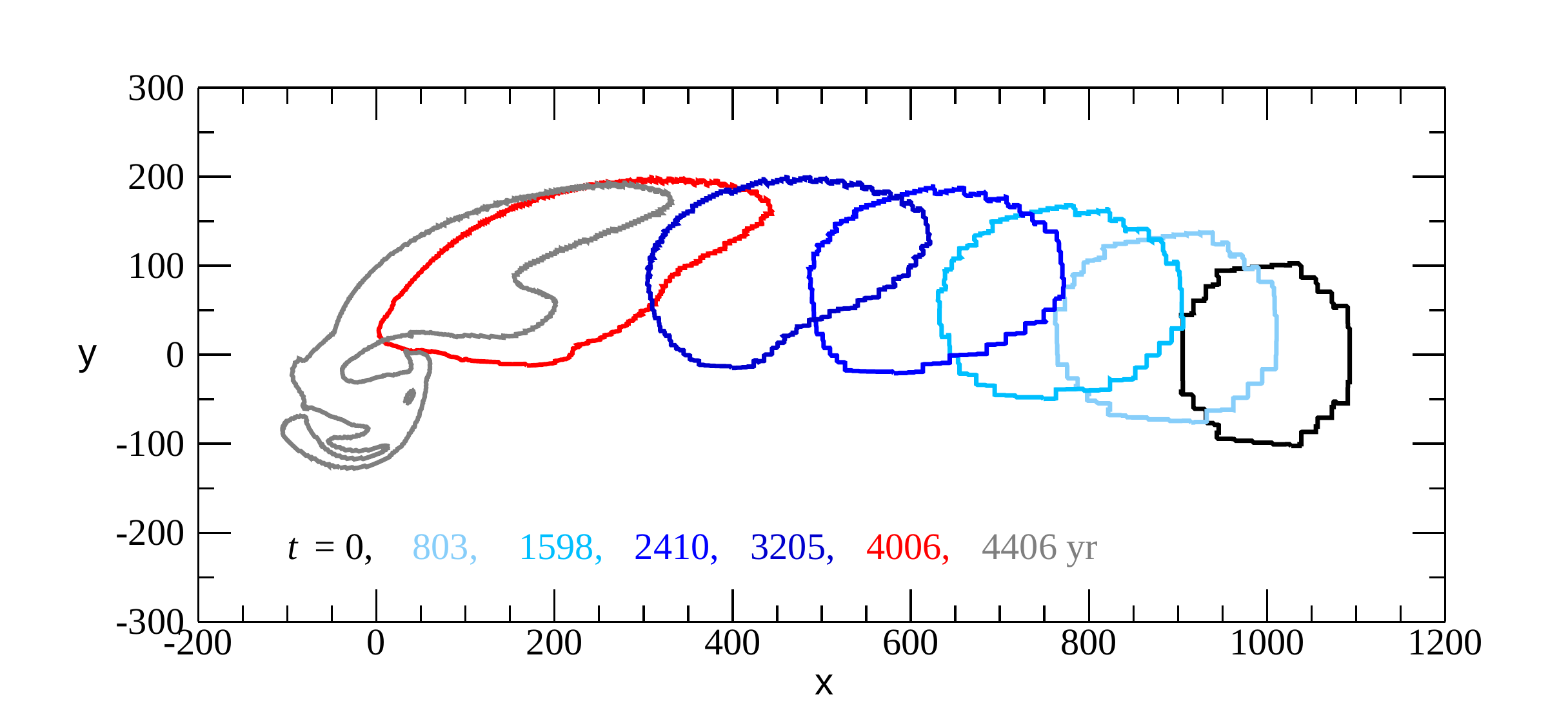}
    \caption{The evolution of the newly accreting
    cloudlet in model A$^\prime$ (from 1,000 au away). }
    \label{fig:projectedAd}
\end{figure}

Model A$^\prime$ has the same initial condition as that of model A
except for the initial distance of the cloudlet to the star.
The initial distance is set to be $ r _0 $ = 1000 au while
it is $ r _0 $ = 500 au in model A.  The initial velocity
of the cloudlet is reduced to 0.97~km~s$^{-1}$. Figure~\ref{fig:projectedAd}
is the same as Figure~\ref{fig:projected} but for model A$^\prime$.
The cloudlet changes its form during the flight approaching to
the star.  The cloudlet has a fin-like structure on the side close to the star before
the collision. The deformation is apparently larger on the rear side.
The deformation is mainly due to the interaction with the warm gas,
though it may be partly due to the
tidal force. The head of the cloud is decelerated by the ram pressure,
while the rear side is not.

\begin{figure*}
\plottwo{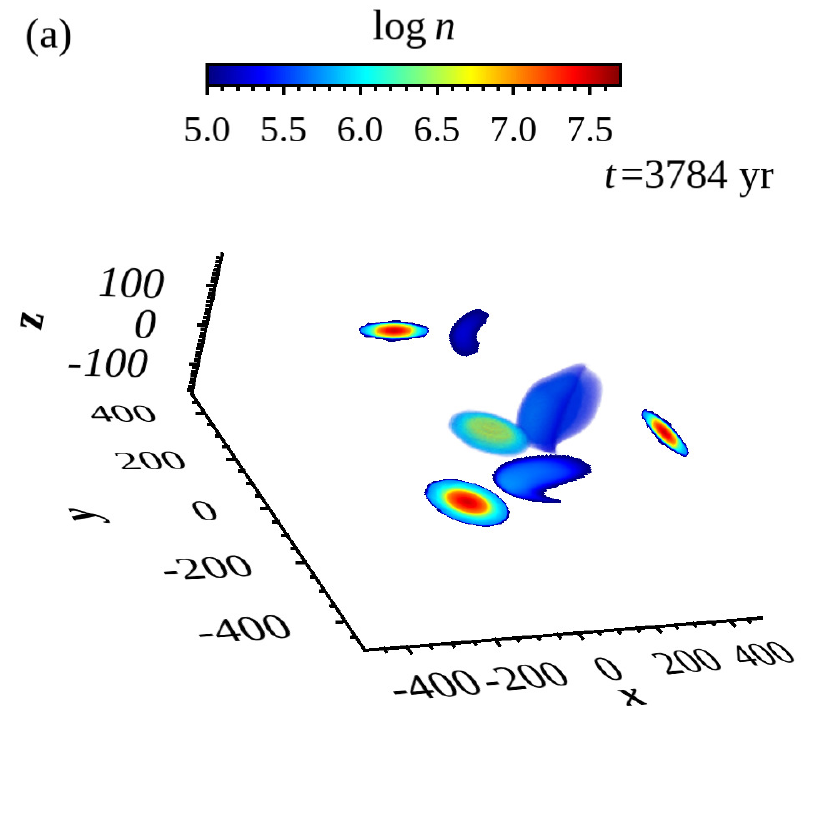}{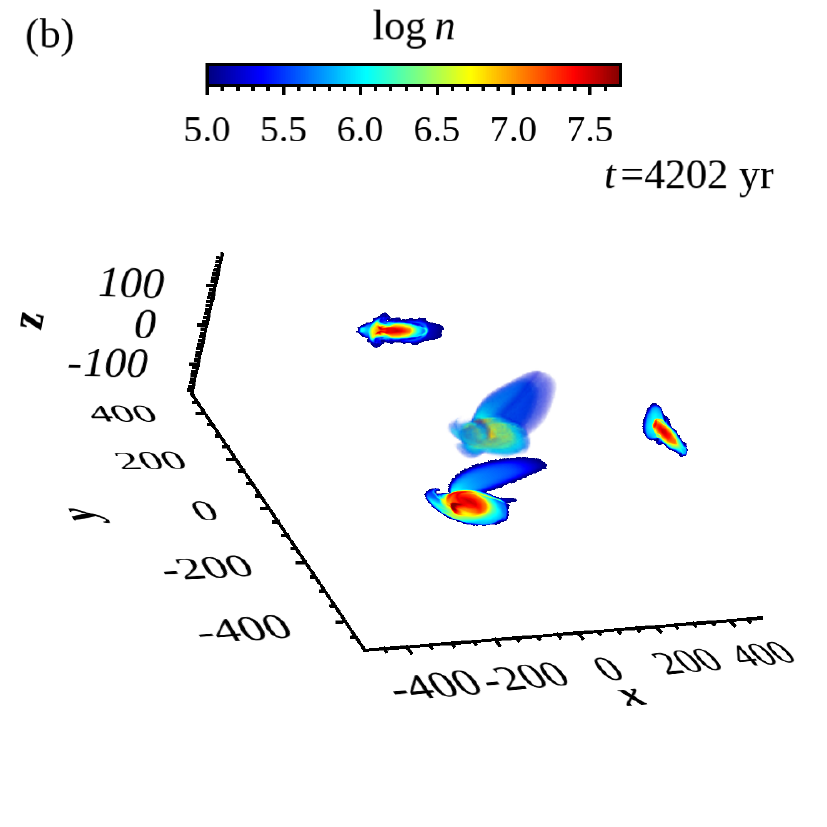}
\caption{The same as Fig.~\ref{fig-init} but for model A$^\prime$ 
at $t = 3784 $~yr and 4202 yr. \label{modelAD}}
\end{figure*}

Figure~\ref{modelAD} shows the cloudlet and disk in model A$^\prime$ at 
$ t = 3784 $ and 4202  yr.  The notaion is the same as that of Figure~\ref{fig-init}.  
The left panel shows a stage before the collision, while
the right panels shows and an early stage of the collision. 
The cloudlet collides with the disk around $ t \simeq 3900 $~yr. 
The delay of the collision is ascribed to the initial distance.
As shown
in Figure~\ref{modelAD}a, the cloudlet is deformed appreciably 
in model A$^\prime$. But the difference is minor as seen from comparison
between  Figures~\ref{fig-init}b
and \ref{modelAD}b. The collision is so violent that it erases subtle differences
in the cloudlet.  We also note that our model of a cloudlet is highly ideal.
A real cloudlet is unlikely to be a uniform sphere at any distance.
Thus we conclude that the assumed initial distance of the cloudlet 
is not a critical parameter.

\subsection{Model B}

We assumed in models A and A$^\prime$ that the orbital plane of the cloudlet coincides with that
of the disk.  However, a cloudlet may approach to the protostar from above
the disk plane.
We have constructed model B to examine the case in which
the orbital plane of the cloudlet is inclined to the disk plane.
The inclination angle is set to be $ \psi _c = 30 ^\circ $.
The other model parameters are the same as those of model A.

\begin{figure*}
\plottwo{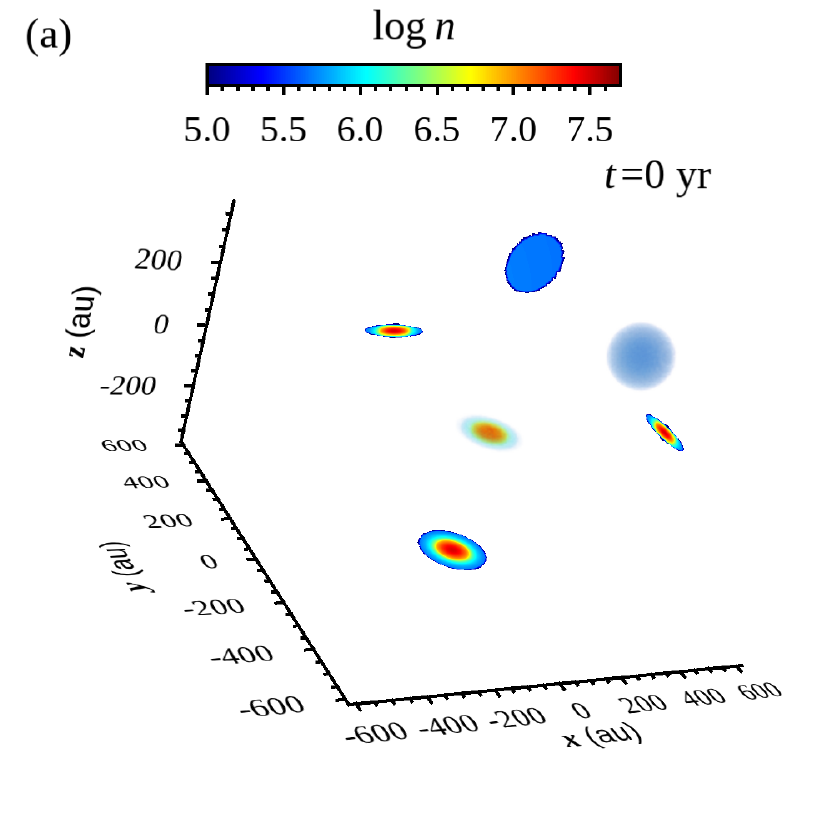}{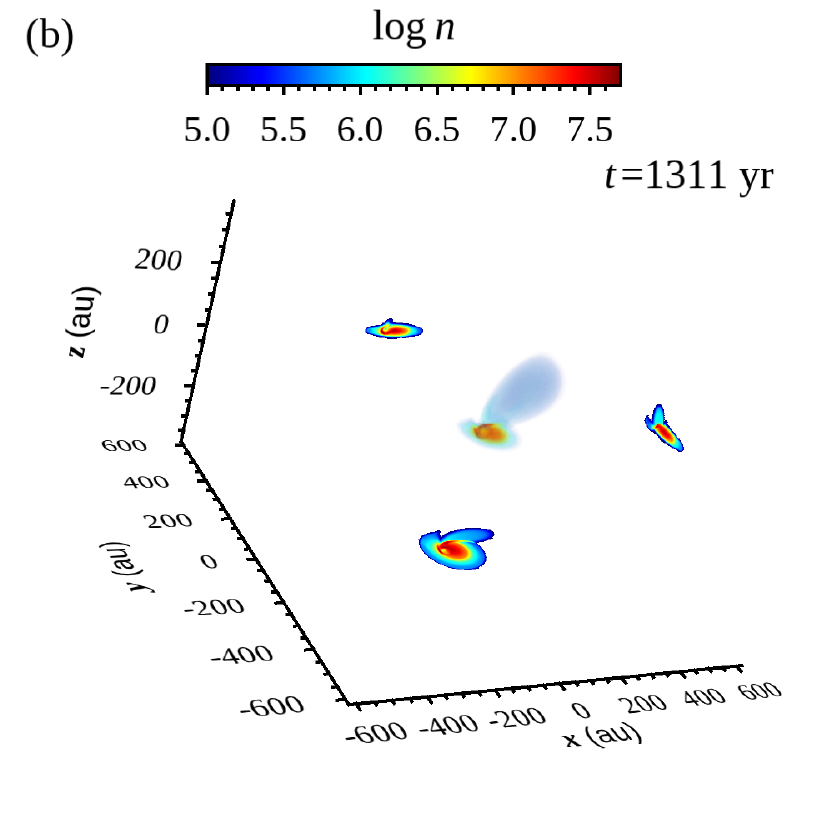}
\caption{The left panel shows the initial gas distribution in model B
while the right panel denotes that at $ t = 1311 $~yr.
The animation shows the evolution of the density distribution
in model B from $ t = 0 $~yr to 2831~yr.\label{B-init}}
\end{figure*}

Figure~\ref{B-init} is the same as Figure~\ref{fig-init} but for
model B at the initial stage ($ t = 0 $) and an early stage
of collision ($ t = 1311~\mbox{yr}$), while the enclosed animation shows the
time evolution. The inclination of the
orbit should not have a siginificant effect on the cloudlet before
the collsion, since the warm gas is assumed to be spherically symmetric.
The inclination changes the geometry of the collision.
The cloudlet collides with the disk from the upper surface 
in model B, though it does from the outer edge in model A.

\begin{figure*}
\plottwo{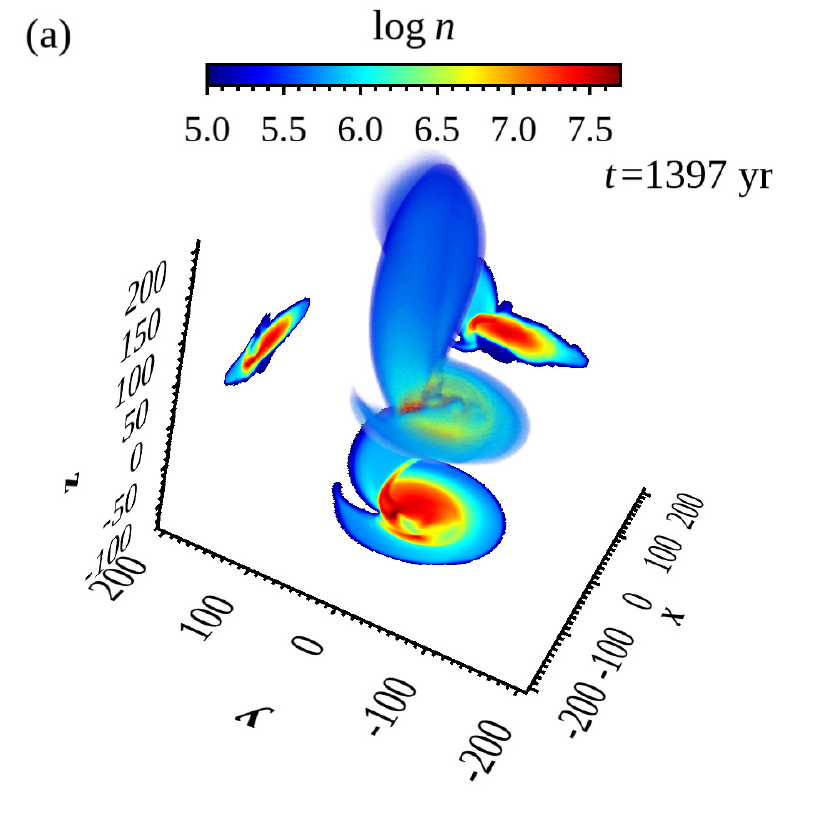}{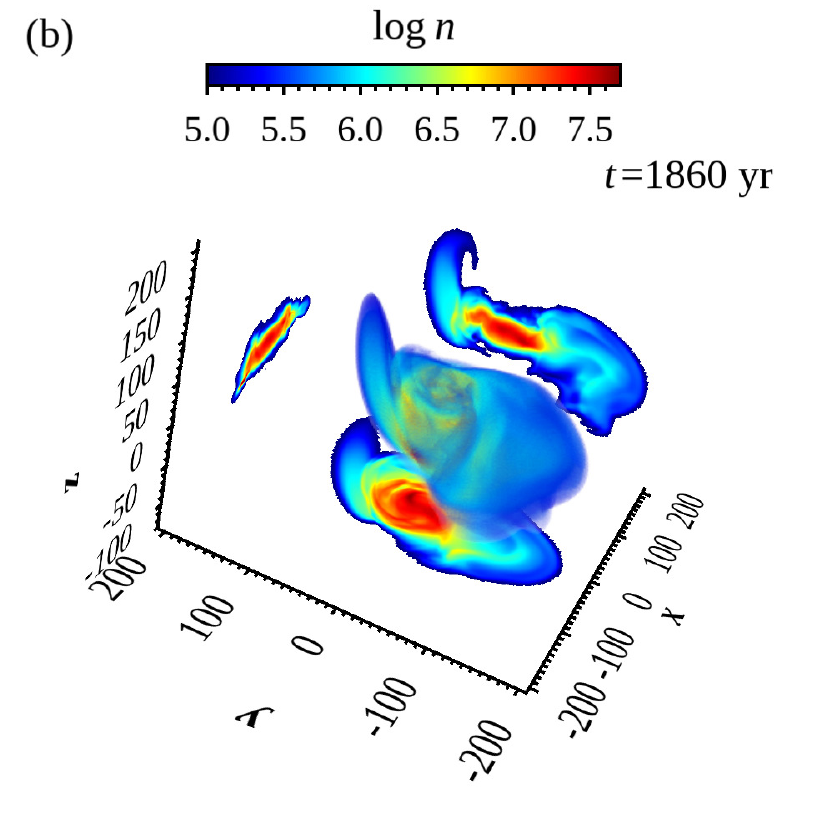}
\caption{The gas distribution around the protostar in
model B at $ t = 1397 $~yr (left) and 1860~yr (right).
\label{B-zoom}}
\end{figure*}

Figure~\ref{B-zoom} gives a zoom-in view of model B at $ t = 1397~\mbox{yr} $
and $ t =1860~\mbox{yr}$.  The cloudlet is stretched by the tidal force to have
a tail. The collision of the cloudlet induces a spiral wave in the disk.
The impact is cleary seen in the left panel of Figure~\ref{B-zoom}.
At the same time, the density increases by the strike and a fraction of the disk is shaved.
The cloudlet breaks the disk and goes through the disk midplane.  
The periastron of the cloulet is located under the midplane. 
The cloudlet gas returns to the upper side of disk after the passage of
periastron.  The right panel shows the collision of the cloudlet
with the disk at a later stage.

\begin{figure*}
\plottwo{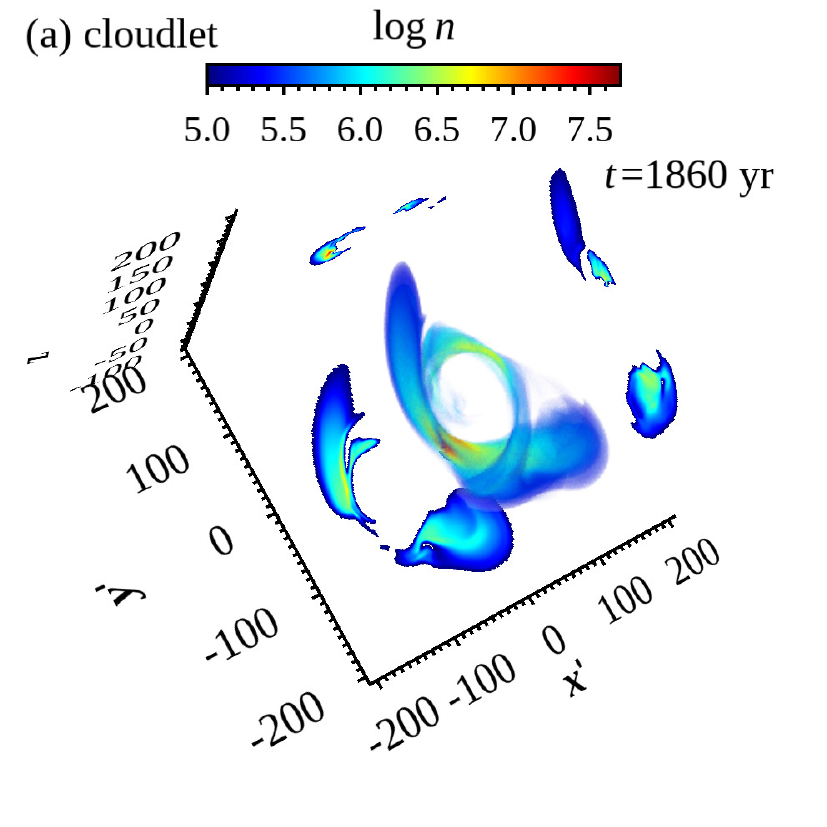}{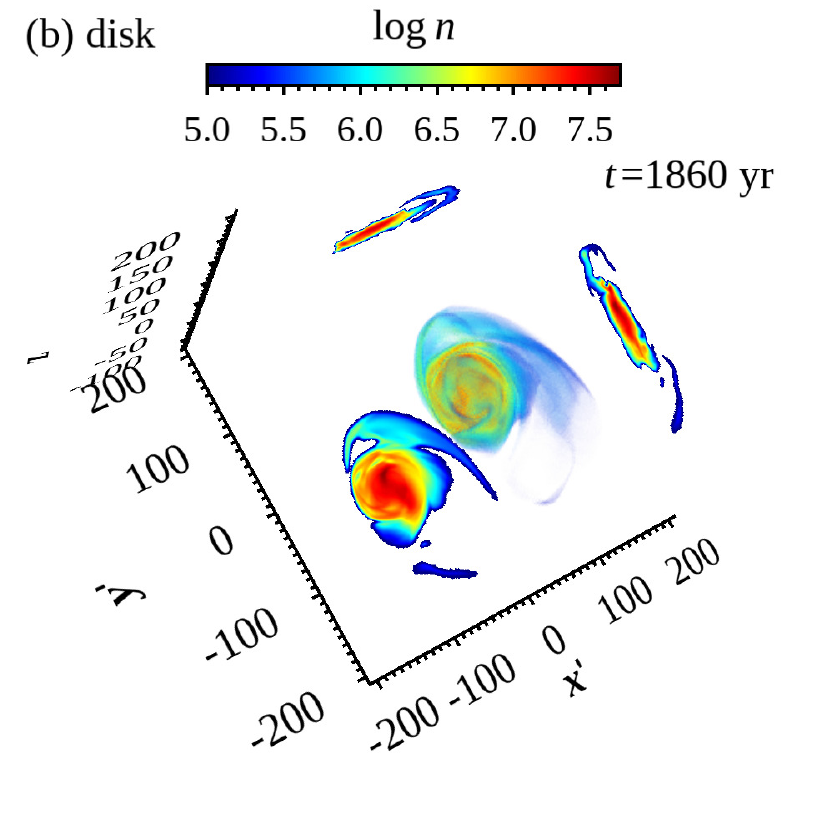}
\caption{The left and right panels denote the cloudlet and
disk gases in model B at $ t = 1860~\mbox{yr}$, 
respectively. The same stage is shown in the right panel
of Fig.~\ref{B-zoom} but from a different viewing angle.
The animation shows the time evolution up to $ t = 2381$~yr.
\label{EDB}}
\end{figure*}

Figure~\ref{EDB} shows the clouldlet and disk at $ t = 1860~\mbox{yr}$ 
separately while the enclosed animation shows the time evolution. Though the same stage is shown in the right panel of Figure~\ref{B-zoom},
the viewing angle and coordinates are different.  Figure~\ref{EDB} employs the
Cartesian coordinates, $ (x ^\prime, y ^\prime, z )$, where
\begin{eqnarray}
 x ^\prime & = &x \cos 37.^\circ 5  + y\sin 37.^\circ 5  ,\\
 y ^\prime & = &- x \sin 37.^\circ 5  + y \cos 37.^\circ 5 ,
\end{eqnarray}
to avoid degeneracy of the axes on the figure.
Thus, the cross sections denote the gas distribution in the
plane of $ x ^\prime = 0 $ and $ y ^\prime = 0 $.
The viewing angle is specified by $ i = 55 ^\circ $, $ \varphi _{\rm obs} = 50 ^\circ $,
and $ \chi _{\rm obs} = 20 ^\circ $.  We use the same viewing angle in the Mock
observation shown in \S 4.  As shown in the panel, gas is ejected
from the disk by the collision with the cloudlet.

\subsection{Models C and D}

We have constructed model C to examine the effects of the initial cloudlet size. The model parameters of Model C are the same as those of model B, except for the initial cloudlet size. The initial cloudlet radius is set to be $ a _c = 200 $~au in model C, while it is 100 au in model B. 
The mass of the cloudlet is $ 1.02 \times 10 ^{-4}~\mbox{M} _\odot$ and is 8 times
higher than that in models A and B.

\begin{figure*}
\plottwo{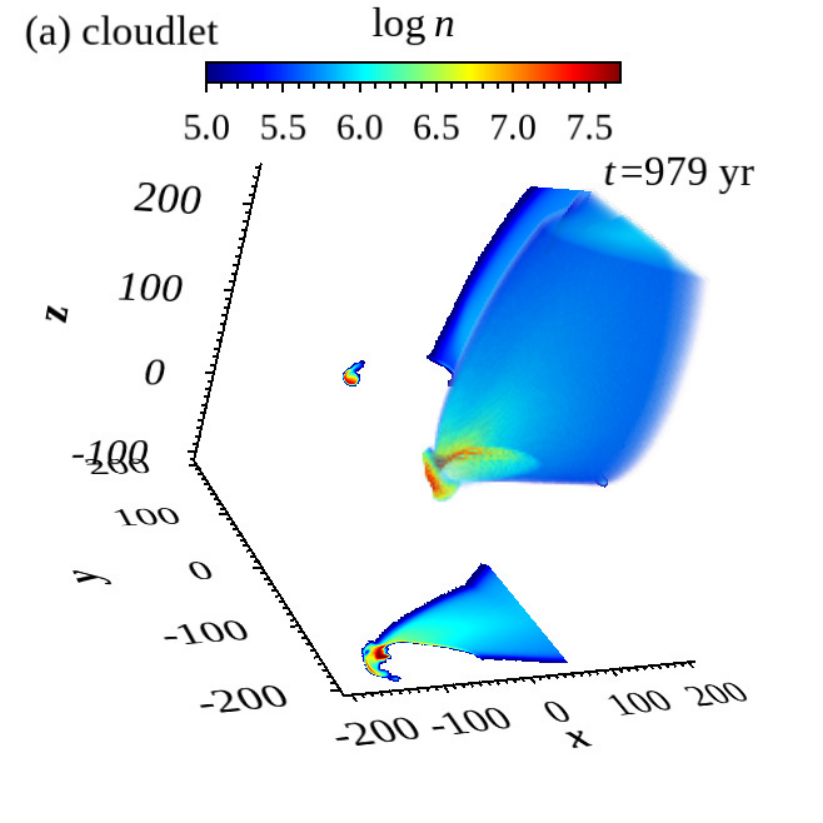}{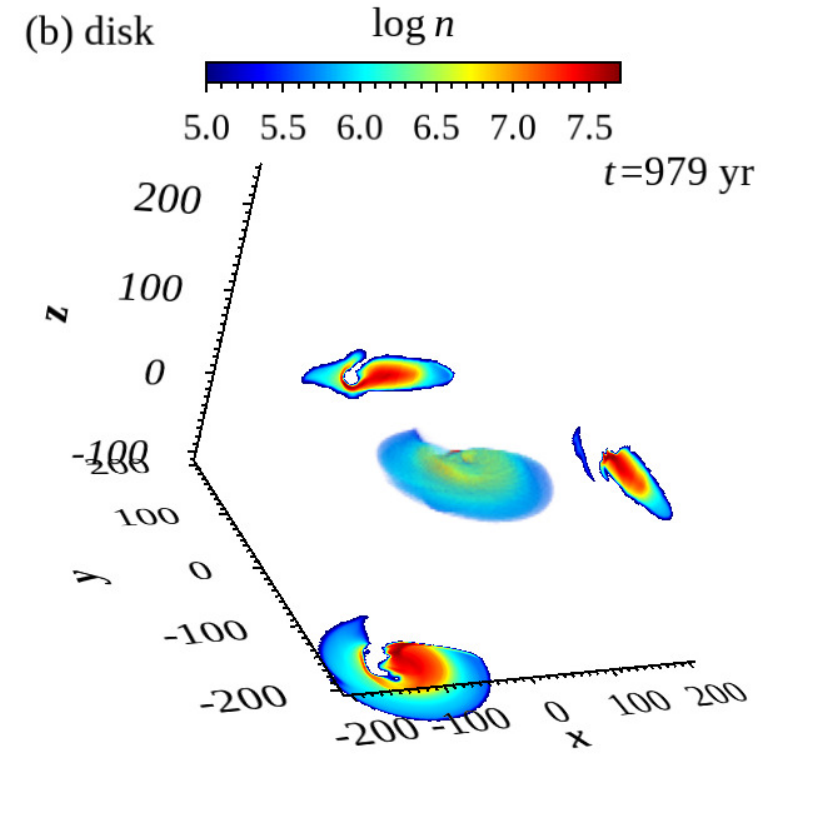}
\plottwo{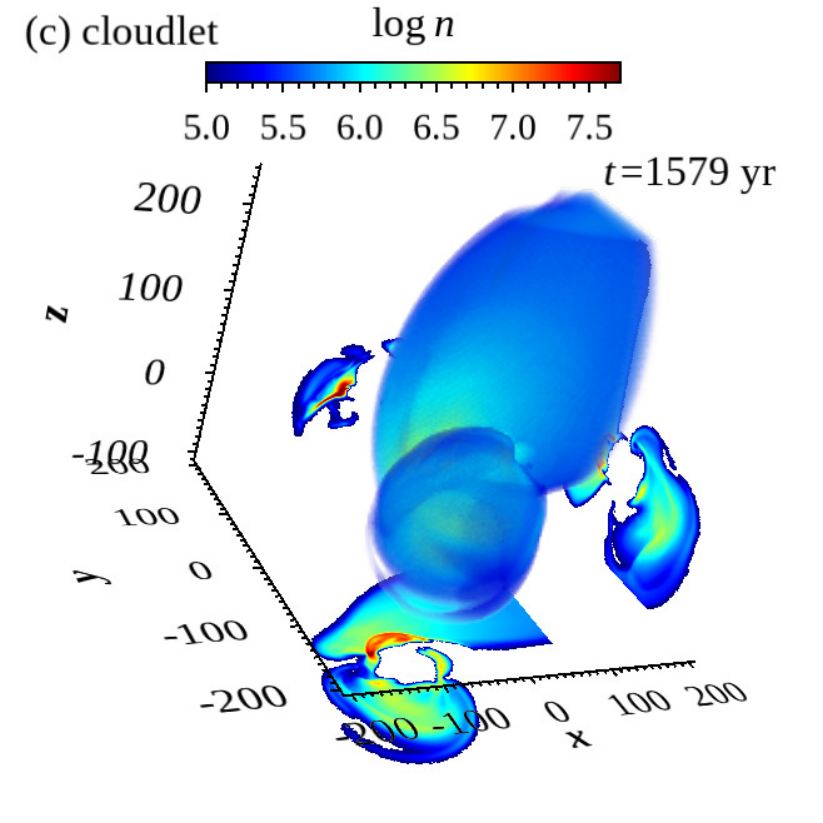}{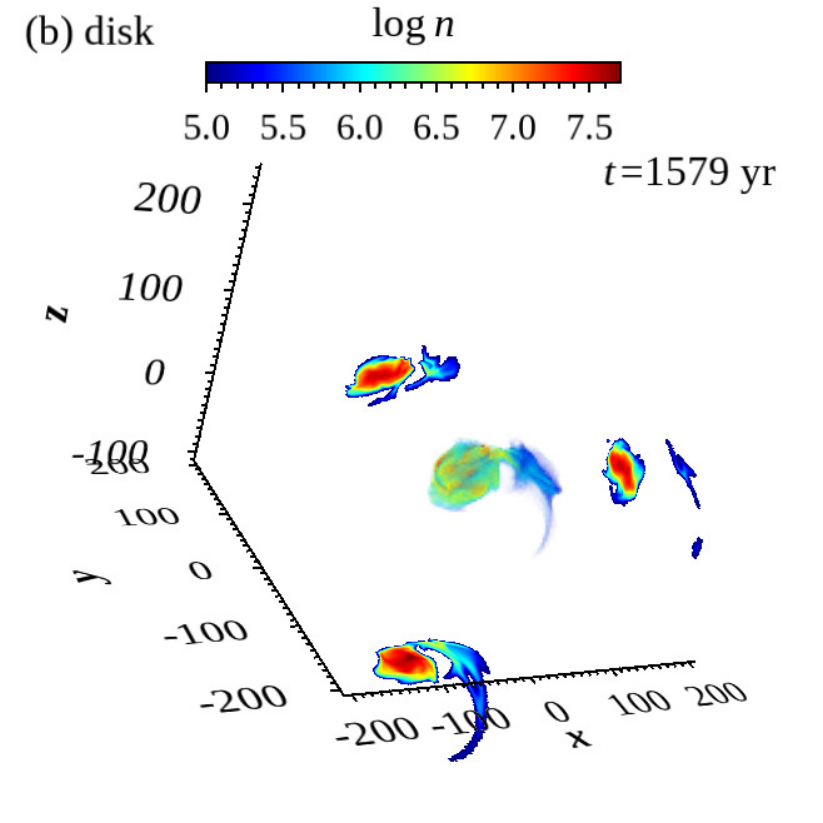}
\caption{Early stage of the collision in model C.  Panels (a) and (c) denote the cloudlet, while panels (b) and (d) do the disk. \label{modelC}}
\end{figure*}

Figure~\ref{modelC} shows two stages after the collision
of the cloudlet with the disk in model C.  The upper and lower panels
show the stages at $ t = 979 $ and 1579 yr, respectively.  The left panels 
denote the cloudlet by the voulme rendering and cross sections, while the right
panels do the disk by the same manner.  
Since the cloudlet is larger,
it reaches the disk a little earlier and occupies a larger volume.
However, the collision forms a narrow arc of high density on the interface
between the cloudlet and disk.  The high density indicates the shock compression
of the cloudlet.  Since the arc is much narrower than the cloudlet, the width of 
the arm is irrelevant to the initial cloudlet size and related to the shock 
strength and cooling efficiency.  Since the gas in the arc is compressed by the shock,
the temperature should increase at once.  In our model, the density is enhanced 
by a factor of $ (\gamma + 1)/(\gamma -1) = 41 $ in the limit of strong shock.

Figure \ref{modelC}b displays that the disk has a slit corresponding to the arc.
The disk outside the slit evolves into an arm at $ t = 1579 $~yr as shown in 
Figure \ref{modelC}d. The inner part of the disk is less affected by the collision.
A significant fraction of the cloudlet is also not affected by the collision.

We have constructed model D to assess the effects of the cloudlet size
on the deformation before the collision.  The parameters of model D are the
same as model C but for $ \psi _{\rm c} $. The orbit of cloudlet is coplanar
to the disk in model D ($ \psi _{\rm c} = 0 $) while it is inclined in
model C.  This difference is insignificant before the collision with disk.
Note that the warm gas is spherically symmetric and the gas disk is disturbed little
before the collision.

\begin{figure}
\plotone{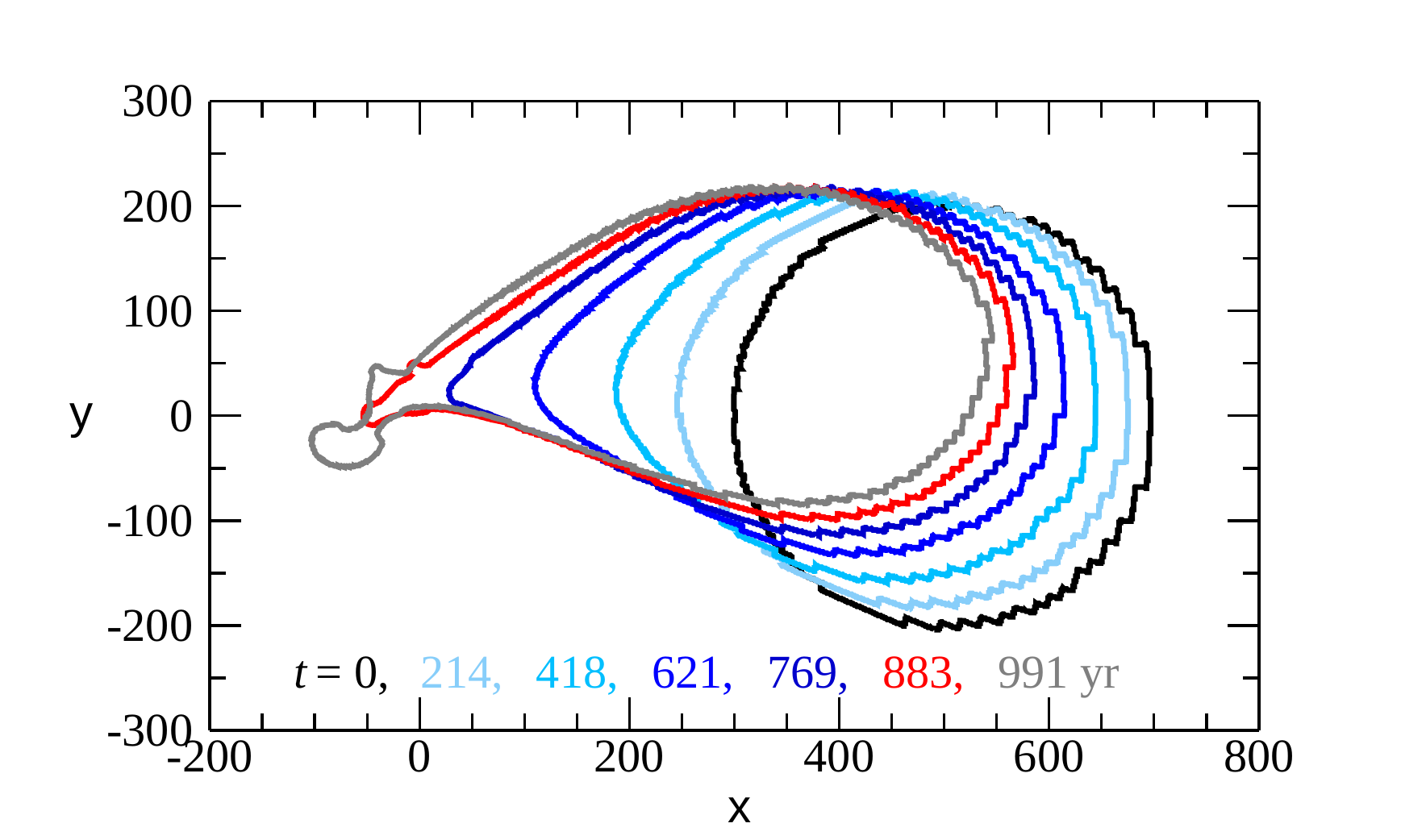}
\caption{The same as Fig.~\ref{fig:projected} but for model D \label{fig:projectedD}}
\end{figure}

Figure~\ref{fig:projectedD} shows the shape of cloudlet projected on the plane
of $ z = 0 $ in model D.  The notation is the same as that of Figure~\ref{fig:projected}.
The cloudlet slims and changes its form from sphere to pear.  
The cloudlet has a trunk stretched toward the protostar before the collision.
The deformation is due to the tidal force and warm neutral gas surrounding the
cloudlet.  Note that the pressure of the warm neutral gas is higher at a shorter
distance from the protostar. The high pressure compresses the cloudlet when it
approaches to the protostar.  The cloudlet changes from a sphere to a pear-like shape before the collision also in model C. So, the cloudlet gives an impact of small
scale on the disk even when it is initially as large as the disk.

The cloudlet begins to cover the whole disk around $ t \simeq 1400~\mbox{yr}$
in model D, though the major part of the cloudlet is still approaching to the 
protostar.  The cloudlet turns around the protostar while disturbing the disk.
Around $ t \simeq 2000~\mbox{yr}$, the cloudlet gas approaching to the protostar
is comparable with that leaving. 

\section{Comparison with Observation}

\begin{figure*}
    \centering
\includegraphics[width=0.9\textwidth]{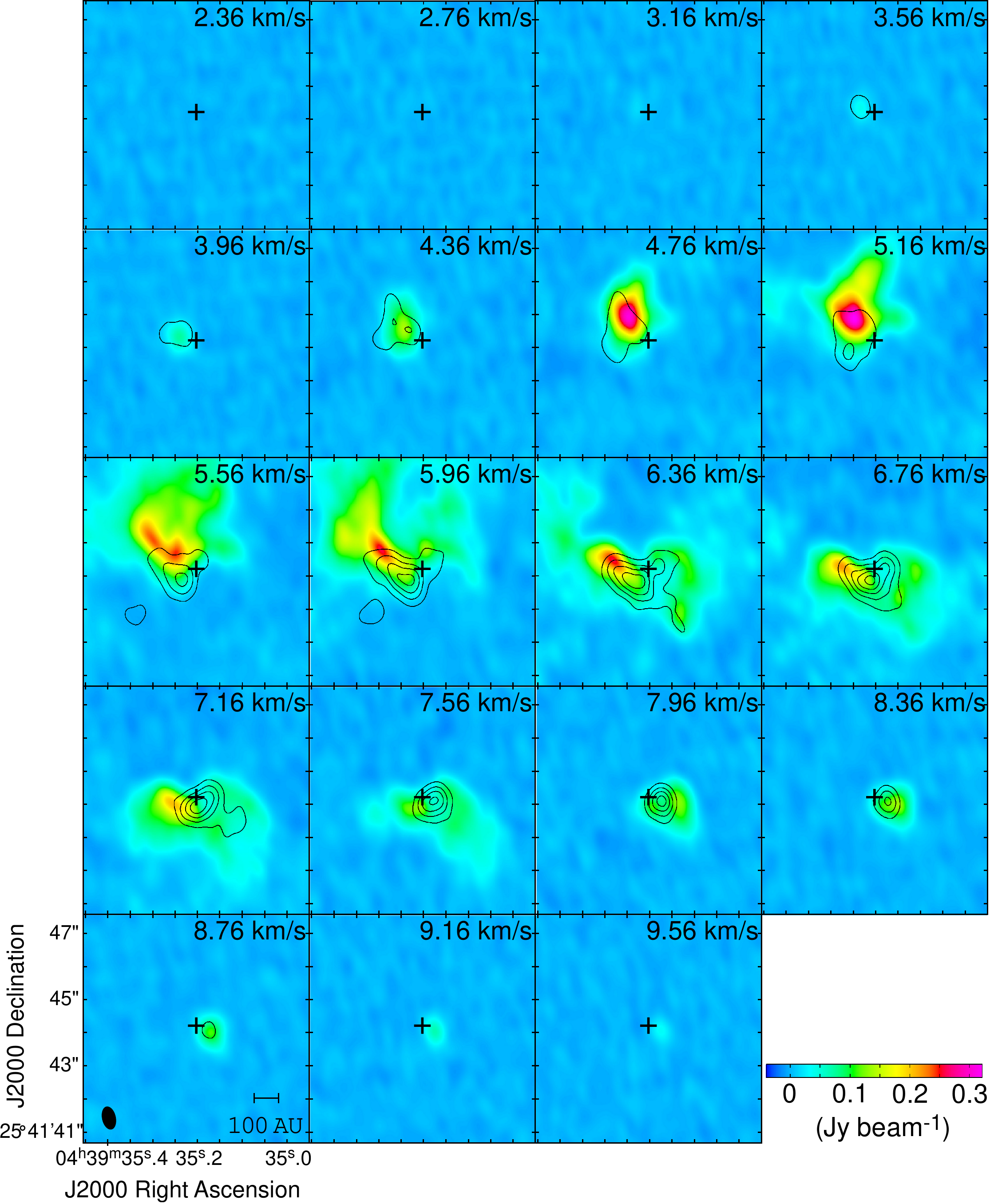}
    \caption{Velocity channel maps of the CS $J$=5-4 line taken by
    ADS/JAO.ALMA\#2013.0.01102.S \citep[see details in][]{sakai16}. 
    The lowest contour
and the contour interval are 50 mJy.   The cross marks represent the position of the continuum peak.}
    \label{fig:CS}
\end{figure*}

In this section we compare our model with the observation of
TMC-1A.  Figure~\ref{fig:CS} shows the line emission of 
CS ($J$ = 5-4; 244.9355565 GHz $E_{\rm u}$ = 35 K) and SO ($J _N = 7_6 - 6 _5$  261.8437210 GHz, $E _u$ =48 K) by the
channel maps.  The data were taken by ALMA (ADS/JAO.ALMA\#2013.0.01102.S, PI:N.Sakai), which were analyzed and reported by \cite{sakai16}.
The beam size is $0{\farcs}69 \times 0{\farcs} 40~(\mbox{PA} = 11^\circ)$
and accordingly 97~au~$\times$~56~au on the sky plane at 
the distance of $ 141 \pm 7 $~pc \citep{zucker19}.  The beam size and linear
scale are shown in the upper left channel map 
of $ V = 2.36~\mbox{km~s} ^{-1}$. Each channel map covers the area of 
$ \Delta\alpha = 6.^{\prime\prime}87$ in the right ascension and 
$ \Delta \delta = 7.^{\prime\prime}00 $ in the declination. The crosses denote
the continuum peak, $ (\alpha _{2000}, \delta _{2000}) = (04^{\rm h}39^{\rm m}35.2,
25^\circ 41^\prime 44.^{\prime\prime}19$.
The color denotes the intensity of CS while
the contours do that of SO.  The color scale is
given in the right bottom corner, while the contour denote
$ I _\nu = $  50, 100, 150, 200, and 250 mJy~beam$^{-1}$.   
We assume the systemic velocity of TMC-1A to be 
$6.36~\mbox{km~s}^{-1} $ according to the channel maps, which is slightly lower than the reported value,
6.6~km~s$^{-1}$, \citep{yen13,harsono14} although the difference ($\sim 0.24~\mbox{km~s}^{-1}$) is comparable to the thermal linewidth.
The blue-shifted emission is much stronger than the red-shifted one \citep[See also figures 1b and 1c of][]{sakai16}.
The CS emission has a strong peak in the region north-east to the protostar. 
We find diffuse emission in the channel maps of $ V = 5.56 $ and 
$ 5.96~\mbox{km~s}^{-1} $, while not in
those of $ V = 6.36 $ and $ 6.76~\mbox{km~s}^{-1} $. 

The SO emission is strong in the region south of the protostar. This region corresponds to 
the impact of the cloudlet to the disk in our model, 
as mentioned in \S\S 3.1 and 3.4. This feature of the SO emission is similar to the 
DG Tau and HL Tau cases observed by \cite{garufi22}.  The foot point of the streamer on the disk is
bright in DG Tau and HL Tau. The line of sight velocity matches with the estimate based
on the kinematical model.  

\begin{figure*}
\plotone{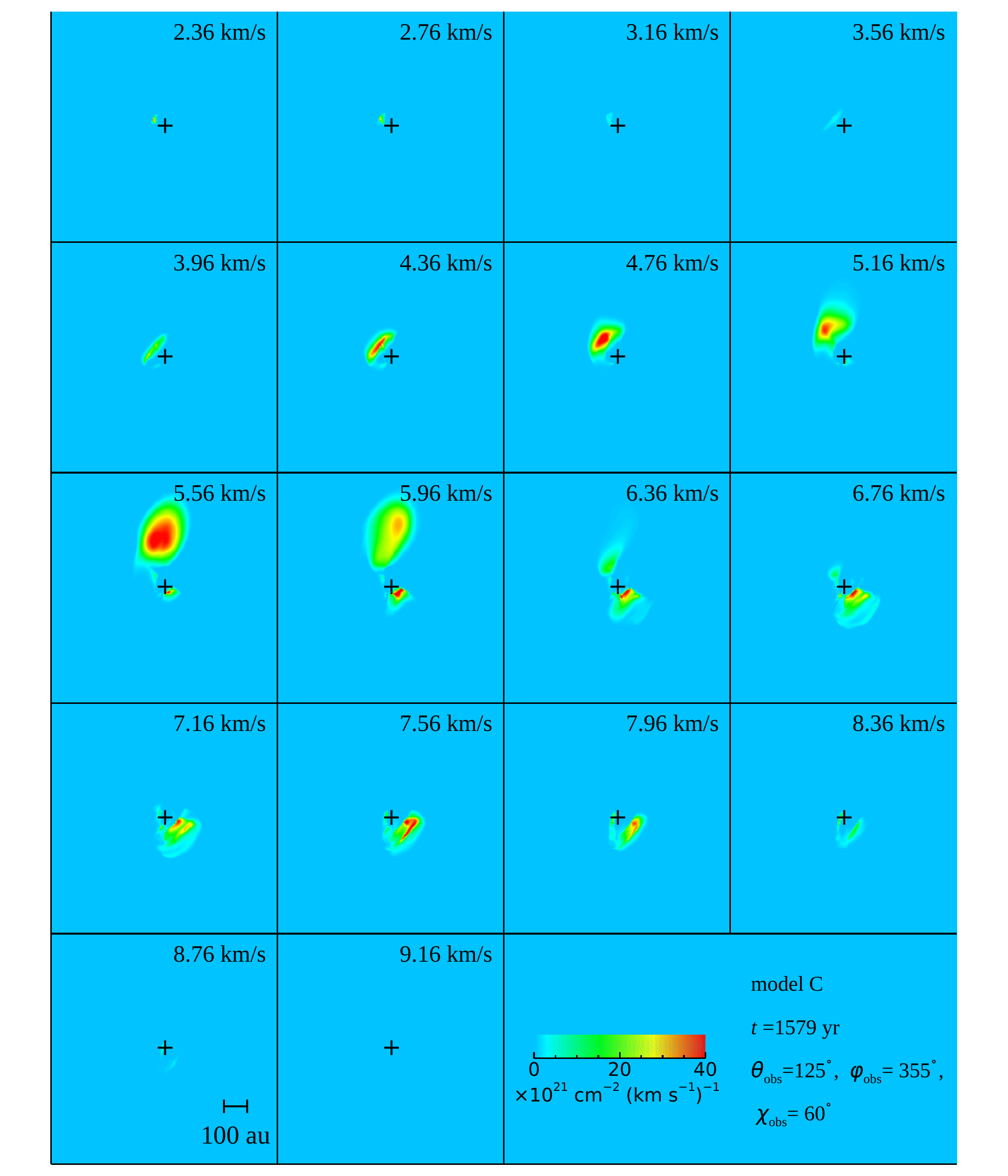}
    \caption{Mock channel maps of model C  at $ t = 1860~\mbox{yr}$.
    The color denotes the density integrated over the line of sight, i.e.,
    $\Sigma (X, Y, V)$ given by Eq. (\ref{sigma-los}).
    The observer's line of sight is specifed by $\theta _{\rm obs} = 125 ^\circ $
    ($ i = 55 ^\circ$), $\varphi _{\rm obs} = 355 ^\circ$, and
    $\chi _{\rm obs} = 60 ^\circ$. The line of sight velocity increases
    from the top left $V = 4.50 $~km~s$^{-1}$ to the right bottom 
    8.36~km~s$^{-1}$, where the systemic velocity is assumed
    to be 6.40~km~s$^{-1}$. 
    \label{channel-37210355285}}
\end{figure*}

Model C reproduces the observed features qualitatively.
Figure~\ref{channel-37210355285} is mock channel maps made 
based on the pseudo observation of model C at
$ t = $ 1860~yr.
Each panel shows the area 
963~au $\times$ 987~au on the sky, which corresponds to $6.^{\prime\prime}8 \times 7.^{\prime\prime}0$ at the distance of 141~pc. 
The color denotes the column density along the line of sight in 
the specified range of the line of sight velocity.  
The inclination angle is assumed to be $ i = 55^\circ$ \citep[0$^\circ$ for face-on,][]{harsono14}.
The viewing angle is specified by $ \varphi _{\rm obs} = 355 ^\circ $ and
$ \chi _{\rm obs} = 60 ^\circ $.
Since the CS emission traces mainly the newly accreted gas seen in L1527 \citep{sakai14a,sakai14b}, we assume that
only the gas of $ c > 0 $ (i.e., the cloudlet) contains CS in 
making the channel map.  The mock channel maps show an arc in the
range of $ 3.96~\mbox{km~s}^{-1} \le V \le 4.76~\mbox{km~s} ^{-1} $.
This corresponds to the former cloudlet gas compressed by the 
collision with the disk.

\begin{figure*}
\plotone{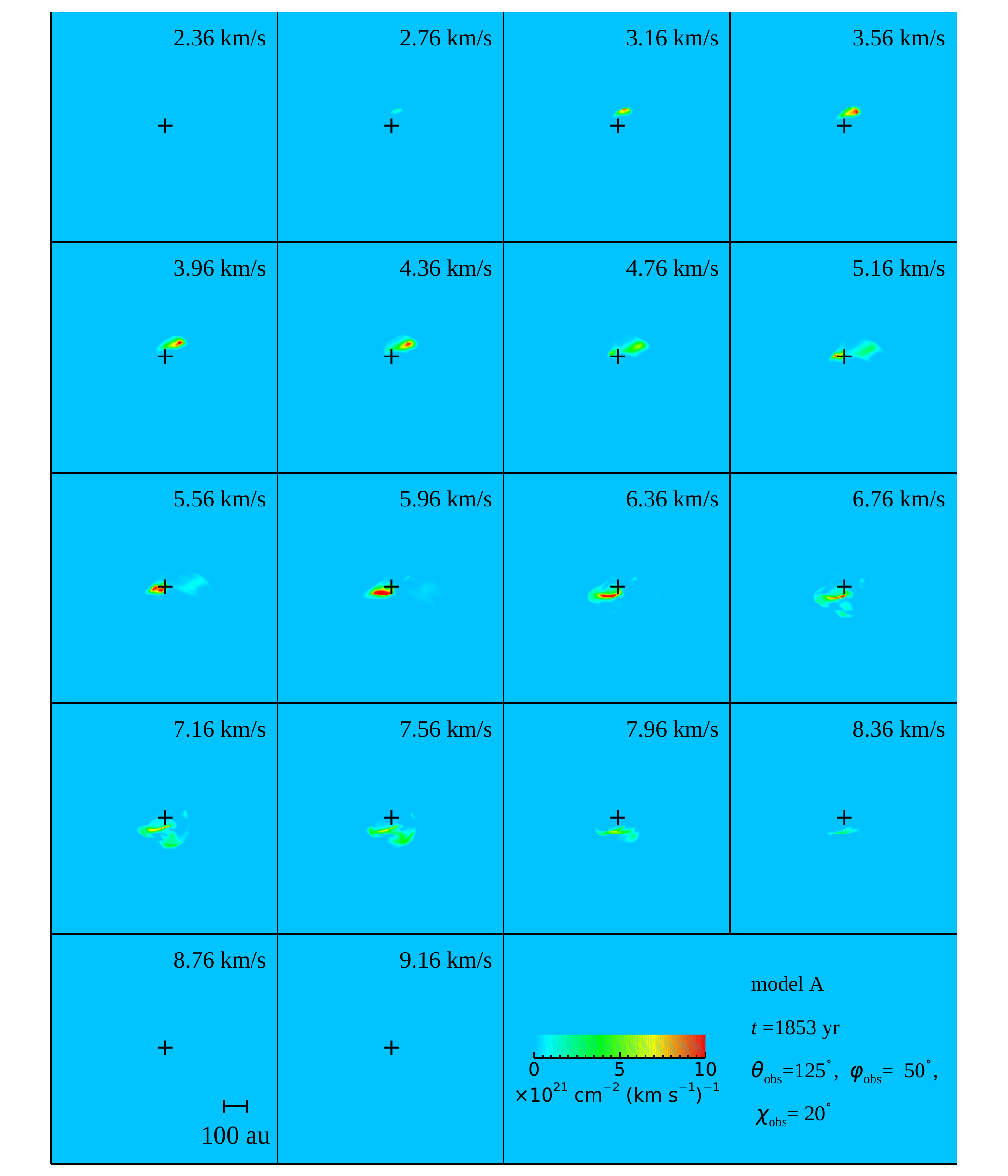}
    \caption{The same as Fig.~\ref{channel-37210355285} but
    for model A at $ t = 1853~\mbox{yr}$.
    {The viewing angle is given by $ \theta _{\rm obs} = 125 ^\circ $
    ($i = 55 ^\circ$)}, $ \varphi _{\rm obs} = 50^\circ $
    and $ \chi _{\rm obs} = 20 ^\circ $
}
    \label{channel-24250050560}
\end{figure*}

Model A cannot reproduce the observed features. 
Figure~\ref{channel-24250050560} is the same as 
Figure~\ref{channel-37210355285} but for model A at
$ t = 1853~\mbox{yr} $. The emission appears
in the channel maps of $ 3.16~\mbox{km~s} ^{-1} \le V \le 
3.96~\mbox{km~s}^{-1} $, though the observation does not
show such highly blue-shifted emission.

The main difference of model B from model A is the orbital
plane of the cloudlet. While the cloudlet has nearly the same 
velocity in both the models, the line of sight velocity depends
on the viewing angle in model C. The orbit of cloudlet is inclined
by $ 85 ^\circ $ for the viewing angle of Figure~\ref{channel-37210355285},
while it is inclined by $ 55 ^\circ $ irrespectively of the viewing angle in
model A. Note that the Keplerian rotation velocity is 3.07~km~s$^{-1}$ at
$ r = 50~\mbox{au}$.  The cloudlet should have a velocity of
$ \sim 4.4~\mbox{km~s}^{-1}$ at $ r = 50~\mbox{au}$ if it follows
the parabolic orbit.  We can reconcile this high velocity with
the observed relatively low blue-shift only when the orbit
is significantly inclined to the disk.
The line of sight velocity is
reduced by the projection.  Since the orbit is close to face-on
in model C, the line of sight velocity is low.
Model B cannot reproduce the spatial extent  of blue-shifted emission,
though the amount of the Doppler shift can be adjusted by the inclination of
the cloudlet orbit.

\cite{aso15} measured the infall velocity from the C$^{18}$O 
($ J = $ 2-1) emission taken with ALMA.  The measured infall
velocity is only 30~\% of that expected from their free-fall 
model in which the orbital plane of the infalling gas coincides
with the disk.  Though they ascribed the low infall velocity to
deceleration by magnetic force, it may be due to the geometrical
effect. If the C$^{18}$O emitting gas is coplanar with
the cloudlet of model C, the line of sight velocity is much
lower than that expected for it to be coplanar with the disk.

If IRAS 04365+2535, the protostar in TMC-1A, were a close binary,
it could be a source of asymmetry. The SO emission might be associated with
a component of the binary. However, we need another explanation for
the blue-asymmetry of the CS emission for this case.  The localization of the CS emission 
around the protostar indicates that the gas accretion is on a short timescale.
If the accretion were continuous, the molecular emission should be more 
extended.  The asymmetry suggests that the gas infall dominates over the rotation.
If the rotation were dominant, the asymmetry should be erased out by differential
rotation.  The localization and asymmetry favor temporal and asymmetric
gas accretion such as clouldlet capture.

The arc-lie features seen in the channel maps of $ v = 4.36 $ and
4.76~km~s$^{-1}$ may correspond to the spiral-like feature discovered by
\cite{aso21}.  They observed TMC-1A with SMA
and ALMA at $\lambda = 1,3=$mm and discovered a spiral-like feature
in the continuum emission. They derived the spiral by subtracting component
symmetric around the star from the high resolution image of the disk.
The residual after the subtraction appears in in the East side of the disk
and is associated with the blue-shifted C$^{18}$O emission 
in the range $(v = 4.35 -- 5.15~\mbox{km~s}^{-1})$.  
See Figure 7 of \cite{aso21} where the channel map of C$^{18}$O is
overlaid on the residual intensity.

\cite{sakai16} noticed another asymmetry in the line emission of
SO ($J_N = 7_6$-$6_5$). The SO line, which could be used 
to a shock tracer, is found to be stronger in south-west to west part of the protoostar 
\citep[Figure 1 of][]{sakai16}.  This component is also seen in the red shifted 
components of the CS map (7.56-9.16km~s$^{-1}$),  which is also consistent with model C.
Panel (c) of Figure~\ref{EDA} and
the right panel of Figure~\ref{EDB} show the disk at
at $ t =  1853~\mbox{yr}$ in model A and
that at $ t = 1860~\mbox{yr} $, respectively.  Both
the disks have a partial loss in the northern side of the disk.
The loss formed by the collision with the cloudlet rotates
faster than the cloidlet.  The disk rotation assimilates the loss 
and is weaker in model B than in model A.

\section{DISCUSSIONS}

So far, the physical structure of the disk/envelope system of TMC-1A has been discussed by using observations of a single molecule at a time. For instance, \cite{aso15} analyzed the C$^{18}$O data by a combination of the Keplerian model and the infalling model, while \cite{sakai16} analyzed the CS data by the infalling rotating envelope model \citep{sakai14a,oya14}.
\cite{sakai16} also pointed out the weak shock feature around the centrifugal barrier of the infalling rotating envelope. These results gave important information on local physical processes in the complex disk/envelope system, but their origins and mutual relations have not been clarified under a broader picture. 
As shown in the previous sections, our cloudlet capture model
can reasonably explain overall features observed in line emissions of 
CS, C$^{18}$O, and SO.  It validates the picture presented by
\cite{sakai16}.  A cloudlet reaches the centrifugal barrier
and a part of it recedes from the protostar again.  
Since the orbit of the receding gas does not intersect with
the approaching gas, the infall continues unless the pre-existing
disk is a serious obstacle. A cloudlet transforms into an arc or a stream during the infall 
as shown in Figure~\ref{modelC}. Both the head and tail are confined in narrow areas while they
have different velocities. Thus, it can explain why the line emission is confined in a narrow area 
in each channel map. The collision of a cloudlet with disk can also explain 
the asymmetric SO bright spot in the observation. Furthermore, it can explain relatively
slow line of sight velocity if the orbit of cloudlet is nearly face-on.

As stated in \S 1, high asymmetry is seen in some 
young protostars even when they are not close binaries. Their asymmetries may also be explained by
a cloudlet capture. The asymmetric gas distribution around the disk/envelope system is often seen in low-mass protostellar sources. For instance, \cite{yen14} observed the Class I protostar L1489 IRS in the CO and its isotopologue lines with ALMA and revealed the red-shifted gas falling to the red-shifted edge of the large Keplerian disk. Since the corresponding structure is not seen in the blue-shited side, this feature can be regarded as an asymmetric accretion. More recently, \cite{pineda20} studied the Class 0 protostar Per-emb-2, which is a close binary system with separation of 20 au, in the HC$_3$N lines with NOEMA. They found an elongated gas clump with the size of 10500 au streaming to the protostar from one side. This feature is quite similar to the case of TMC-1A.  

Very recently, \cite{garufi22} have reported streamers in DG Tau and HL Tau.
The streamers are visible in the CO and CS emission lines in DG Tau, while in HCO$^+$ and
CS emission lines in HL Tau. They have also detected SO and SO$_2$ line emission
from the foot point of the streamer on the disk. Since SO and SO$_2$ are good tracers of 
a shock, the emission is an evidence that the streamer is a trail of infalling gas.
They confirmed the infalling gas scenario by an analytic streamline model.
Interestingly, they argue that the southern streamer is continuation of the northern
one and hence an outflow in DG Tau \citep[see Fig. 7 of][]{garufi22}. 
Their interpretation supports our model that the
red-shifted component is continuation of the blue-shifted one in TMC-1A.

Though we ignored the magnetic field in our modeling for simplicity,
the cloudlet may be permeated by magnetic field.
If the magnetic field is strong enough, it should decelerate the infall
of the cloudlet appreciably as suggested by \cite{aso15}. 
However, \cite{garufi22} have succeeded in reproducing the streamers
in DG Tau and HL Tau by taking account of the gravity only.
This means that the magnetic field is weak in DG Tau and HL Tau.
It seems reasonable to assume that the magnetic field is also weak in TMC-1A.
Weak magnetic field may play a role in late evolution. 
Gas ejection seen in our simulations would be changed, if
initially weak magnetic field were taken into account.
\cite{unno22} have shown that initially weak magnetic field
is amplified by the collision of cloudlet.  When the cloudlet
is larger than the disk thickness, the amplified magnetic field
accelerate and eject a part of the cloudlet from the system.

The asymmetric feature is also seen around some  other Class II sources. \cite{huang21} observed the CO (2-1) emission toward GM Aur with ALMA and found the blue-shifted gas extending from the disk on a 1000 au scale. They interpret it as the remnant gas of the envelope or the cloud component infalling to the disk.  They also point out that the nearby Class II sources, SU Aur and AB Aur, would have a similar asymmetric feature, on the basis of the Herschel SPIRE data. Above all, the asymmetric feature seems more or less frequent occurrence in young sources. 

It is interesting to examine the possibility that
gas accretion onto a protostar is mainly through
cloudlet capture.  If it is the case, the accretion
rate is highly variable in nature.  The capture of 
cloudlet can change the disk rotation axis which is
parallel to the total angular momentum of the disk.
Each cloudlet should have a different angular momentum
vector and the capture should change the direction.
The change may result in launch of multiple outflows
in IRAS 15398-3359 observed by \cite{okoda21}.
Some cloudlets may recede from the protostar before
reaching a close vicinity of it.  

It should be noted that existence of a warm gas is 
a key issue in our cloudlet capture model. As mentioned
by \cite{dullemond19,kueffmeier19,kueffmeier21}, an isothermal cloudlet
expands and disperses if it is not confined by pressure.
A theory of protoplanetary disk also invokes a surrounding
warm tenuous gas \citep[see, e.g.,][]{dutrey14}. The disk surface should also be 
in pressure equilibrium with the warm gas.  Then,the cloudlet
and disk surface have nearly the same density since they have
the same pressure and nearly the same temperature. If the 
density of the cloudlet is nearly the same as that of the disk,
the impact of the cloudlet collision should be significant
since the ram pressure exceeds the gas pressure of the disk.
The disturbance by the collision will induce shock heating and 
mixing of disk gas, both of which will affect the chemical evolution
of the disk.

We point out that the inclination of the orbital plane of the cloudlet 
to the disk may enhance the shock strength.  When the cloudlet is coplaner,
the radial velocity vanishes at the centrifugal barrier. The slow radial
velocity may cause a weak shock.  However, the cloudlet has a still large 
azimuthal velocity.  If the orbital plane is inclined, a fraction of it 
results in a relative velocity with the disk. Since the rotation velocity
reaches several km~s$^{-1}$, a fraction of it is still supersonic and 
will contribute to the shock.  


\begin{acknowledgments}
TH thanks Yosuke Matsumoto for his contribution on 
the visualization of our numerical simulations.
We thank an anonymous reviewer for their constructive comments.
This work was supported by JSPS KAKENHI Grant Numbers
JP18H05222, JP19K03906, JP20H05845, JP20H05847, JP20H00182.
This study used the ALMA data set ADS/JAO.ALMA\#2013.0.01102.S.
ALMA is a partnership of the European Southern Observatory, the National Science Foundation (USA), the National Institutes of Natural Science (Japan), the National Research Council (Canada), the NSC and ASIAA (Taiwan), and KASI
(Republic of Korea), in cooperation with Republic of Chile. The Joint ALMA Observatory is operated by ESO, the AUI/
NRAO, and NAOJ.
\end{acknowledgments}

\facility{ALMA}






\end{document}